\documentclass[a4paper,11pt]{article}

\usepackage{cite}

\usepackage{graphicx}
\usepackage{amssymb,flushend}

\newcommand{\RR}{\mbox{${\rm \:  R\!\!\!\! I
\;\;}$}}

\newcommand{\vs}{\vspace{0.25cm}}

\newtheorem{theorem}{Theorem}
\newtheorem{itlemma}{Lemma}[section]
\newtheorem{itproposition}[itlemma]{Proposition}
\newtheorem{itcorollary}[itlemma]{Corollary}
\newtheorem{itremark}[itlemma]{Remark}
\newtheorem{itremarks}[itlemma]{Remarks}
\newtheorem{itdefinition}[itlemma]{Definition}
\newtheorem{itexample}[itlemma]{Example}

\newenvironment{lemma}{\begin{itlemma}\rm}{\end{itlemma}} %no-italics
\newenvironment{remark}{\begin{itremark}\rm}{\end{itremark}} %no-italics
\newenvironment{remarks}{\begin{itremarks} \rm}{\end{itremarks}}
\newenvironment{corollary}{\begin{itcorollary}\rm}{\end{itcorollary}}
\newenvironment{proposition}{\begin{itproposition}\rm}{\end{itproposition}}
\newenvironment{definition}{\begin{itdefinition}\rm}{\end{itdefinition}}
\newenvironment{example}{\begin{itexample}\rm}{\end{itexample}}
\newenvironment{fact}{\noindent {\em Fact}. \ \ }{\hfill \medskip}
\newenvironment{proof}{\noindent {\em Proof}.\ \
}{\hspace*{\fill}$\Box$\medskip}
\newenvironment{claim}{\noindent {\em Claim}. \ \ }{\hfill \medskip}

\newcommand{\be}[1]{\begin{equation}\label{#1}}
\newcommand{\ee}{\end{equation}}
\newcommand{\bl}[1]{\begin{lemma}\label{#1}}
\newcommand{\br}[1]{\begin{remark}\label{#1}}
\newcommand{\brs}[1]{\begin{remarks}\label{#1}}
\newcommand{\bt}[1]{\begin{theorem}\label{#1}}
\newcommand{\bd}[1]{\begin{definition}\label{#1}}
\newcommand{\bp}[1]{\begin{proposition}\label{#1}}
\newcommand{\bc}[1]{\begin{corollary}\label{#1}}
\newcommand{\bfact}[1]{\begin{fact}\label{#1}}
\newcommand{\bex}[1]{\begin{example}\label{#1}}
\newcommand{\ec}{\end{corollary}}
\newcommand{\efact}{\end{fact}}
\newcommand{\eex}{\end{example}}
\newcommand{\el}{\end{lemma}}
\newcommand{\er}{\end{remark}}
\newcommand{\ers}{\end{remarks}}
\newcommand{\et}{\end{theorem}}
\newcommand{\ed}{\end{definition}}
\newcommand{\ep}{\end{proposition}}
\newcommand{\epr}{\end{proof}}
\newcommand{\bpr}{\begin{proof}}
\newcommand{\bcl}{\begin{claim}}
\newcommand{\ecl}{\end{claim}}

\newcommand{\bi}{\begin{itemize}}
\newcommand{\ei}{\end{itemize}}
\newcommand{\ben}{\begin{enumerate}}
\newcommand{\een}{\end{enumerate}}
\newcommand{\text}[1]{\hbox{\rm \ #1\ \/}}

\title{Indirect Controllability of Quantum Systems;
A Study of Two Interacting Quantum Bits}

\author{Domenico D'Alessandro
 and Raffaele Romano
 \thanks{Manuscript received February 15, 2011. D. D. research was supported by NSF under Grant No.
 ECCS0824085. R. R. acknowledges financial support from the Parisi Foundation. This research is partially
 supported by the ARO MURI grant W911NF-11-1-0268.}
 \thanks{D.D. is with the Department of
 Mathematics, Iowa State
 University, 440 Carver Hall, Ames IA-5001, Iowa,
 U.S.A.; e-mail: daless@iastate.edu.}
 \thanks{R.R. is with the Department of Physics, Universit\`a degli Studi
 di Trieste, Strada Costiera 11, I-34151  Trieste, Italy; e-mail: rromano@ts.infn.it.}}

\begin{document}

\maketitle

\begin{abstract}

A quantum mechanical system $S$ is indirectly controlled when the
control affects an ancillary  system $A$ and the evolution of $S$ is
modified through the interaction with $A$ only. A study of
 indirect controllability gives a description
 of the set of states that can be
obtained for $S$ with this scheme.
In this paper, we study    the indirect controllability of quantum
systems in the finite dimensional case. After discussing the
relevant definitions, we give a general necessary condition for
controllability in Lie algebraic terms. We present a detailed
treatment of the case where both systems, $S$ and $A$, are
two-dimensional (qubits). In particular,  we characterize the
dynamical Lie algebra associated with $S$+$A$, extending previous
results, and prove  that complete controllability of
$S$+$A$ and an appropriate notion of indirect controllability are
equivalent properties for this system. We also prove several further
indirect controllability properties for the system of two qubits,
and illustrate the role of the Lie algebraic analysis in the
study of reachable states.

% and give example of constructive algorithms for indirect controllability

\end{abstract}

\vs

\vs

\vs

%\begin{keywords}
%Indirect Control of Quantum Systems,
%Interacting  Quantum Systems, Lie Theoretic  Methods
%\end{keywords}

\vs

\vs

%http://www.cs.bris.ac.uk/~montanar/presentations/QuantumWalks.ppt
%%%%%%%%%%%%%%%%%%%%%%%%%%%%%%%%%%%%%%%%%%%%%%%%%%%%%%%%%%%%%%%%%%%%%%%%%%%%%%%%

\section{INTRODUCTION}

Indirect controllability of a quantum system $S$ refers to the
situation where two quantum systems $S$ and $A$ interact, and the
externally applied control is allowed to influence the evolution of
$A$ only, while the state of $S$ is of interest. The systems $S$ and
$A$ are referred to as the {\it target system} and {\it accessor
system}, or {ancillary system}, respectively. Therefore, the state of $S$ is
{\it indirectly} controlled through the interaction  with
$A$. The problem of {\it indirect controllability} is to analyze to
what extent it is possible to modify the state of $S$ with this
scheme in various situations.
%In general, one would like to give algebraic
%characterizations to describe this property.

The question of indirect controllability is of theoretical interest
for the further development of quantum control and of quantum
information, as indirect controllability is a signature of the
capability of the dynamics to generate {\it entanglement} \cite{NC}.
In fact, if there is no entanglement between the two
systems,  $S$ and $A$, they evolve separately and indirect
controllability is impossible. This question is also of practical
relevance since, in many experimental setups, it is much easier to
control the accessor system $A$ rather than the relevant part $S$.
There might be several reasons for that. For instance, we might be
able to use only control fields tuned  at the resonance frequency of
some of the particles which make up a multi-particle systems or we
might be able to access only systems  in certain locations. As a
physical example of indirect control, protocols of short-distance
quantum communication have been described using {\it spin
chains} \cite{bose, caneva}. The state to be transmitted is put in
interaction with one end of the chain, the controlled part, and it
is reproduced at the other side of the chain, without physical
transmission. So, control at one end of the chain realizes the task
of state transfer. It has been proved theoretically \cite{Daniel}
that control at the end of the chain is sufficient to realize a
reliable transfer. This can be seen as indirect control. The target
system $S$ is the last spin at one end of the chain whose state we
want to manipulate (and make equal to the one at the other end). The
rest of the spin system can be seen as the accessor system. Spin
chain schemes can be implemented through super-conducting devices
based on Josephson junctions \cite{you} and are potential tools
for the implementation of quantum technologies.

%%%%%%%%%%%%%%%%%%%%%%%%%%%%%%%%%%%%%%%%%%%%%%%%%%%%%%%%%%%%%%%%%
%Possible scenarios
%of indirect control have also been considered in the literature as
%an alternative or complementary strategy to coherent control
%(\cite{noiIncoherent}, \cite{Daniel}, \cite{Fu} and the references
%therein), examining various  physical systems and assumptions from a
%more theoretical perspective.
%%%%%%%%%%%%%%%%%%%%%%%%%%%%%%%%%%%%%%%%%%%%%%%%%%%%%%%%%%%%%%%%%

%The study of controllability of quantum systems has a large
%literature
There is a large literature studying the control of quantum systems,
starting with establishing the connection with geometric
control theory \cite{JS,HT,RSD,Butsam} until
more recent studies for closed (see, e.g., \cite{confraIEEE,altaf,Clarkluc,SchirmerSom1,SchirPull,SchirmerSom2}) and open (see e.g., \cite{alta2,Dirr,Wu111}) quantum systems. Nevertheless, an algebraic
characterization of indirect controllability is still lacking. In
fact, most investigations on indirect controllability have given
conditions so that the full system $S+A$ is {\it completely
controllable}, that is, every unitary transformation can be
performed on it.
%(cf. definitions in section \ref{definizioni}).
This implies in particular that every unitary operation can be
performed on $S$. In these cases,  controllability can be proved by
verifying the standard Lie algebraic conditions for closed systems,
that is,  calculating  the so-called {\it dynamical Lie algebra}
(see definition in Section \ref{general}), which fully characterizes
the dynamics of the total system \cite{Mikobook}. However, as we
shall prove, there are cases where indirect controllability on $S$,
in an appropriate sense, can be achieved even without complete
controllability of the total system, and this motivates a more
detailed study of this property.

%Our previous work \cite{noiIncoherent} has considered indirect controllability through
%the state of $A$ with a constant (without time varying control)
%Hamiltonian for $S$ and $A$, being both  two level systems.

The goal of this paper is to provide a general framework for the
 treatment of indirect controllability of quantum systems, with special
 emphasis on  the case of two coupled qubits $S$ and $A$.
 Because of the introduction of the partial trace (see definition in the next section)  to describe the
dynamics of the system $S$, much of the machinery of Lie
transformation groups cannot be used in this case. Nevertheless Lie
algebraic techniques play a crucial  role in determining
 many indirect controllability properties.

We briefly summarize this work. In Section \ref{matheframe} we
define the mathematical set up and give the basic notations and
mathematical tools. In Section \ref{general} we define the main
notions for indirect controllability and give a general necessary
criterion for indirect controllability at the Lie algebra level
(Theorem \ref{gennegat}).
%We recall  how the state of a quantum system
%is described in quantum mechanics, the dynamics of two coupled
%systems $S$ and $A$ and how the state of the system $S$ is extracted
%through the operation of partial trace (cf.
%(\ref{definitiopartialtrace})). We give the main definitions of
%indirect controllability. This Section also contains a summary of
%the main technical results presented in the rest of the paper.
%In Section \ref{general} we take up the general treatment of indirect
%controllability for finite dimensional quantum systems.
%The dynamics of the total system is determined by a Lie algebra ${\cal L}$, which
%is called the {\it dynamical Lie algebra} which  determines many
%indirect controllability properties. The main tool used in our
%investigation is the theory of Lie algebra and Lie groups. However,
%because of the introduction of the partial trace to describe the
%dynamics of the system $S$, much of the machinery of Lie
%transformation groups cannot be used in this case (cf. Remark
%\ref{Invariance}).
In Sections \ref{2qbit} and \ref{indcon} we present a detailed
treatment of the case of two interacting quantum bits, $S$ and $A$.
In particular, in Section \ref{2qbit} we give a description of the
dynamical Lie algebra associated with  this system. This extends a
result of \cite{Fu} which only gives a sufficient condition for
complete controllability. Based on the characterization of the
dynamical Lie algebra, in Section \ref{indcon} we prove various
indirect controllability properties for the two-qubit system
presented in Theorems \ref{UICequiv}, \ref{CCimpliesFIC}, in
Proposition \ref{pure9} and in Example \ref{contoesempio}. We draw
some conclusions and give directions for further research in Section
\ref{conclusioni}.

% and this more general
% approach we will take in this paper. We shall, in fact,  give
% examples where indirect controllability is verified without complete
% controllability.

\section{BASIC FACTS CONCERNING THE DYNAMICS OF INTERACTING SYSTEMS} \label{matheframe}

The states of the target and ancillary systems, $S$ and $A$,  are
described by {\it density operators} $\rho_S$ and $\rho_A$,
respectively, that is, positive semi-definite, trace class
operators\footnote{A trace class operator is a bounded operator $A$
%on a Hilbert space (with inner product $\langle \cdot, \cdot \rangle$)
%such that  for some orthonormal basis $e_k$ $\sum_k \langle
%(A^\dagger  A)^{\frac{1}{2}} e_k,e_k \rangle$ is finite.
such that ${\rm Tr}(A^\dagger  A)^{\frac{1}{2}}$ is finite. We shall deal
with the finite dimensional case where this requirement is
automatically satisfied.} with unit trace on Hilbert spaces ${\cal
S}$ and ${\cal A}$, associated with $S$ and $A$, respectively
\cite{NC,Petruccione}. The total system $S + A$ is described by the
density operator $\rho_{T}$ on the Hilbert space ${\cal S} \otimes
{\cal A}$. The sets of density operators on ${\cal S}$, ${\cal A}$,
and ${\cal S} \otimes {\cal A}$ are convex, the boundary consisting
of {\it pure states}, that is, projectors of rank one of the form
$\rho = |\psi \rangle \langle \psi|$, with $|\psi\rangle$ in ${\cal
S}$, ${\cal A}$, or ${\cal S} \otimes {\cal A}$,
respectively.\footnote{We shall make use of the Dirac notation of
quantum mechanics denoting a general vector in a Hilbert space ({\it
ket})  by $| \cdot \rangle$ and the associated dual linear operator
({\it bra}) by $\langle  \cdot |$.}  The state of a subsystem
($\rho_S$ or $\rho_A$) can be extracted from the state of the total
system $\rho_{T}$ through the partial trace operation, $\rho_S =
{\rm Tr}_A(\rho_{T})$ and $\rho_A = {\rm Tr}_S(\rho_{T})$. We recall that,
if we express an Hermitian operator $\rho \in {\cal S} \otimes {\cal
A}$ as \be{ropt} \rho=\sum_{j} \alpha_j \, \rho^S_j \otimes \rho^A_j,
\ee with Hermitian operators $\rho^S_j \in {\cal S}$ and $\rho^A_j
\in {\cal A}$, and real coefficients $\alpha_j$, the partial trace
operation is a linear map from the space of Hermitian operators on
${\cal S} \otimes {\cal A}$ to the space of Hermitian operators on
${\cal S}$ (${\cal A}$) defined as \cite{NC}
\be{definitiopartialtrace} {\rm Tr}_A(\rho):=\sum_j \alpha_j \, \rho^S_j
{\rm Tr}(\rho^A_j), \, \,  {\rm Tr}_S(\rho):=\sum_j \alpha_j \, \rho^A_j
{\rm Tr}(\rho^S_j).  \ee
%The partial trace represents the correct way to
%extract the state of a subsystem from the state of the embedding
%system. In fact, the resulting state gives the correct statistics
%for any local observable. In particular, for any observable
%(Hermitian operator) $M$ on ${\cal S}$ we have \be{statisticright}
%{\rm Tr}\Bigl(M {\rm Tr}_A(\rho_{T})\Bigr)= {\rm Tr}(M \otimes {\bf 1} \rho_{T}).
%\ee
%(cf. \cite{NC} for further discussion of this point).

The main property of the partial trace which will be used in the
following is that, for unitary operators $F$ and $G$ acting on
${\cal S}$ and ${\cal A}$ respectively, it holds \be{mainpropertyPT}
{\rm Tr}_A\left( F \otimes G \rho F^\dagger \otimes G^\dagger \right) = F \,
{\rm Tr}_A(\rho) F^\dagger. \ee

%For future
%reference, it is useful to denote the space of density operators on
%${\cal S}$ (or ${\cal A}$, ${\cal S} \otimes {\cal A}$) by
%$\Lambda_S$,  $\Lambda_A$ and  $\Lambda_{T}$, respectively.

In our discussion we are going to  assume  {\it finite dimensional}
systems $S$ and $A$, that is,  both ${\cal S}$ and ${\cal A}$ are
finite dimensional vector spaces.
%\footnote{In fact, most of the
%treatment will consider the case where both $S$ and $A$ have
%dimension 2}
Some of the definitions and notions we shall introduce hold for the
infinite dimensional case as well or can be naturally extended. We
shall denote by $n_S$ and $n_A$ the dimensions of ${\cal S}$ and
${\cal A}$, respectively, and the total system $S + A$ has dimension
$n_S n_A$, so $\rho_S$, $\rho_A$ and $\rho_{T}$ are represented by
Hermitian, positive semi-definite, unit trace matrices of dimensions $n_S
\times n_S$, $n_A \times n_A$, $n_S n_A \times n_S n_A$,
respectively.

The evolution of the total system is determined  by a Hermitian
operator $H_{T}$, called the {\it Hamiltonian}, which can be
conveniently separated in four parts as follows:\footnote{Without
loss of generality we shall assume in the following that all
Hamiltonians have zero trace,
%and this is done without loss of
%generality
since we can always decompose a Hamiltonian $H$  as $H =
\frac{1}{n} \, {\rm Tr}(H) {\bf 1}_n + \tilde H $ where $\tilde H$ has zero
trace and $n$ is the dimension of the underlying system. The term
$\frac{1}{n} \, {\rm Tr}(H) {\bf 1}_n$ affects the dynamics only for a global
phase factor which is irrelevant in quantum mechanics, and thus can
be neglected. Here and in the following ${\bf 1}$ denotes the
identity operator. We shall write ${\bf 1}_n$ if the identity
operator acts on a space of dimension $n$, when we want to highlight
such dimension.} \be{jh} H_{T} = H_S + H_A + H_I + H_C. \ee The
Hamiltonian $H_S$ determines the evolution of the system $S$ alone.
It is of the form $H_S = B\otimes {\bf 1}_{n_A}$, for some Hermitian
operator $B$ on ${\cal S}$. Analogously $H_A = {\bf 1}_{n_S}
\otimes {C}$ for some Hermitian operator $C$ on ${\cal A}$. $H_I$
models the two-body interaction between $S$ and $A$, and it has the
general form \be{HIgenform} H_I = \sum_{j=1}^m B_j \otimes C_j, \ee
for Hermitian operators $B_j$ and $C_j$ on ${\cal S}$ and ${\cal
A}$ respectively. Finally the Hamiltonian $H_C$ models the
interaction with the external control. Since in the indirect
controllability scheme we assume only control on $A$, $H_C$ will be
written as \be{formaHC} H_C = \sum_{k = 1}^n u_k(t) {\bf 1}_{n_S} \otimes L_k, \ee
for some control functions $u_k(t)$ and Hermitian
operators $L_k$.\footnote{ We assume linearity in the control as in
dipole interaction, but this assumption is not crucial, and  the
treatment that will follow can be adapted to more general situations with minor modifications.}
Assuming units such that $\hbar = 1$, the total system $S + A$
evolves according to the Schr\"odinger operator equation
\be{Sropeq111} \dot X = - i H_{T} \bigl(u(t)\bigr)X, \qquad
X(0)={\bf 1}_{n_S n_A}, \ee and the density operator $\rho_{T}$
varies as \be{densopev} \rho_{T} (t) = X(t) \rho_{T} (0)
X^{\dagger} (t). \ee The initial state is assumed to be
uncorrelated, that is $\rho_{T} (0) = \rho_S (0) \otimes \rho_A
(0)$. By combining this evolution with the partial trace
(\ref{definitiopartialtrace}), we obtain the dynamics of the
relevant system $S$, \be{evosys} \rho_S(t)= {\rm Tr}_A \left(X(t)
\rho_S(0) \otimes \rho_A(0) X^\dagger (t) \right).  \ee In the
following, we shall study the map, for given $\rho_A$, \be{mappa}
\rho_S \rightarrow {\rm Tr}_A( X \rho_S \otimes \rho_A X^\dagger),  \ee as
$X$ varies in the set of all unitary operators available as
solutions of (\ref{Sropeq111}) with suitable control actions.

\section{INDIRECT CONTROLLABILITY FOR GENERAL SYSTEMS}
\label{general}

%We give several, physically motivated definitions for indirect
%controllability and discuss the relation between the various
%notions. We also set up more definitions and describe how the theory
%of Lie algebras can be used in the investigation of indirect
%controllability.
We now give  several definitions of controllability, that might be
appropriate in different contexts. The following definition is
standard for the  controllability of the composite system $S+A$.

\bd{complete} The system $S+A$ is called {\it completely
controllable} (CC)  if, for any special unitary evolution $P \in
SU(n_S n_A)$,\footnote{We shall use the notation  $\mathfrak{u}(n)$
($\mathfrak{su}(n)$) to denote the Lie algebra of $n \times n$
skew-Hermitian matrices  ($n \times n$ skew-Hermitian matrices with
zero trace)  while $U(n)$ ($SU(n)$) denotes the corresponding Lie
group of $n \times n $ unitary matrices (unitary matrices with
determinant equal to $1$).} there exists a time $T\geq 0 $ and a set
of control functions, $u_1 (t),\ldots, u_m (t)$, defined in $[0,T]$,
such that the solution of (\ref{Sropeq111}) satisfies $X (T) = P$.
\ed

In dealing with indirect controllability of $S$, it is important to
take into account two processes: (i) the preparation of $\rho_A$;
(ii) the dynamical control of the full system $S+A$.
%It turns out
%that different notions of indirect controllability are possible
%depending on how much freedom we assume in the choice of $\rho_A$.
%The notions of indirect controllability of $S$ depend on how much
%freedom we assume in the choice on the initial state $\rho_A$, and
%how much freedom we require in steering $\rho_S$.
The process of {\it preparation} of $\rho_A$ is typically an
irreversible process completely separated from the unitary {\it
evolution} of the total system. This preparation is performed before
dynamically controlling the system. Therefore, in the definition
 of indirect controllability we need to include the information on
 the set of possible initial states for the system $A$. This set is denoted by $\Lambda_A$
 in the following definitions.
 %Accordingly, we find convenient to define
 %the notion of reachable set from an initial state $\rho_S$ as follows.

\bd{reacha} A state $\rho_S'$ is said to be reachable from $\rho_S$
given $\Lambda_A$,  if there exists $\rho_A \in \Lambda_A$, and a
set of controls $u_1 (t),\ldots,u_m (t)$, defined in $[0,T]$, such
that the solution $X = X(T)$ of (\ref{Sropeq111}) satisfies
\be{reachi} \rho_S' = {\rm Tr}_{A} \left(X(T) \rho_S \otimes \rho_A
X^\dagger(T)\right). \ee If ${\cal R}$ is the set of unitary evolutions
which can be obtained for the total system, the  set of reachable
states from $\rho_S$ given $\Lambda_A$ is  \be{reachi2} \Lambda_S
(\rho_S) :=  \{ X \rho_S \otimes \rho_A X^\dagger \vert \rho_A \in
\Lambda_A, X\in {\cal R} \}. \ee \ed

In the previous definition, $\Lambda_S (\rho_S)$ depends on
$\Lambda_A$, and the choice of $\rho_A$ in  $\Lambda_A$ is part of
our control strategy.
%If the reachable set given
%$\Lambda_A$ is  $\Lambda_S(\rho_S)$, we are certain that we can
%choose a state $\rho_A$ and a control law to steer the state of $S$
%to any value in $\Lambda_S(\rho_S)$.
Special cases are when $\Lambda_A$ consists of a single state, that
is the accessor admits only a specific initial configuration, or
when it consists of states that have special features. The case of
pure states is particularly important, both from an application
viewpoint, and because of some results we shall prove in this paper
(cf. proposition \ref{pure9}). Nonetheless, we will not a priori put
any restriction  on $\Lambda_A$.

%%%%%%%%%%%%%%%%%%%%%%%%%%%%%%%%%%%%%%%%%
%Usually, controllability can be defined making reference to the
%allowed state transitions (state controllability), or to the
%operations that can be performed (operator controllability).
%Accordingly, we find convenient to give the following definitions of
%indirect controllability.
%%%%%%%%%%%%%%%%%%%%%%%%%%%%%%%%%%%%%%%%

%Special cases are given in the following definitions.

\bd{FICUIC} The system $S$ is called {\it fully indirectly
controllable} (FIC) given $\Lambda_A$ if, for every initial state
$\rho_S$, the reachable set $\Lambda_S (\rho_S)$ is the set of all
density matrices of $S$. The system $S$ is called {\it unitary
indirectly controllable} (UIC) given $\Lambda_A$ if, for any  given
initial state $\rho_S$, the reachable set $\Lambda_S (\rho_S)$ is
the set of all density matrices unitarily equivalent to $\rho_S$.
\ed

%UIC is the property we would seek to obtain if we were to control $S$
%directly and $S$ was a closed system non interacting with any other
%system.

By definition, FIC implies UIC given the same set $\Lambda_A$.
Moreover, if FIC (or UIC) holds given $\Lambda_A$, it also holds
given any $\Lambda_A^{\prime}$ such that $\Lambda_A \subseteq
\Lambda_A^{\prime}$. It is a known result in quantum information
theory that, if the system is CC and $n_A \geqslant n_S^2$,
the system $S$ is FIC given the set $\Lambda_A$
consisting of one pure  state for $A$.\footnote{This is equivalent
to $\Lambda_A$ being the set of all pure states since under the CC
condition we can transfer the state of $A$ between two arbitrary
pure states. So this statement is equivalent to pure FIC.  It is
proved that an arbitrary completely positive map can be performed on
$S$ (dilation theorem for completely positive maps~\cite{NC}) and,
according to~\cite{Wu111}, every possible state transformation can
be realized through a completely positive map.} We shall prove in
Theorem \ref{CCimpliesFIC} that if $n_S=2$, $n_A=2$
is sufficient to have this property.

%%%%%%%%%%%%%%%%%%%%%%%%%%%%%%%%%%%%%%%%%%%%%%%%%%%%%%
%%%%%%%%%%%%%%%%%%%%%%%%%%%%%%%%%%%%%%%%%%%%%%%%%%%%%%

If UIC or FIC hold independently of the set $\Lambda_A$, we shall
refer to these properties as {\it strong}. For instance, strong UIC
means that, independently of the initial state $\rho_A$, for all
$X_S \in SU(n_S)$ we are able to find a control strategy such that
the solution of (\ref{evosys}) is $X_S \rho_S X_S^\dagger$. Complete
controllability of the total system implies the  strong UIC notion
since every transformation of the form $X_S \otimes {\bf 1}$ is, in
particular, a unitary transformation, and gives the desired state
transfer $\rho_S \rightarrow X_S \rho_S X_S^\dagger$, independently
of the initial state $\rho_A$. For this reason, most of the
investigations on indirect controllability have focused on giving
conditions for complete controllability (see, e.g., \cite{Fu}). When
UIC or FIC holds for the set $\Lambda_A$ of  pure states for $A$ we
shall refer to these properties as {\it pure}. Clearly strong
implies pure for both properties.

\vs

%In the definition \ref{reacha} and the ones that follow, we could
%have restricted ourselves to the case where $\rho_S$ is only allowed
%to be a {\it pure state}. This case is particularly important for
%applications. We shall not introduce a formal definition in order
%not to burden the reader with too many definitions; when necessary,
%we shall discuss in the paper how the results modify or specialize
%if we only consider pure initial states for $S$.

\vs

\br{Invariance} We remark that all the notions of indirect
controllability which have been introduced, and the criteria which
will be presented in the following, are invariant with respect to
unitary changes of coordinates which affect the system $S$ and $A$
separately (local transformations). They are not invariant under
general unitary changes of coordinates that affect both systems $S$
and $A$. \er

%or, equivalently, similarity transformations on the
%Hamiltonians of the form $H \rightarrow X \otimes Y H X^\dagger
%\otimes Y^\dagger$. They are not however invariant under general
%changes of coordinates on the total system as it happens for
%controllability notions in direct control schemes. Moreover, if $X$
%belongs to a group (as we shall see it is the case) the map
%(\ref{mappa}) does not in general describe an {\it action} in the
%sense of Lie transformation groups \cite{LTG} as in general
%\be{noaction} {\rm Tr}_A\left[X_2 ({\rm Tr}_A(X_1 \rho_S \otimes \rho_A
%X_1^\dagger))\otimes \rho_A X_2^\dagger \right] \not={\rm Tr}_A\left[
% X_2 X_1 \rho_S \otimes \rho_A X_1^\dagger X_2^\dagger\right].
%\ee This means that much of the tools of transformation groups which
%were used for example in \cite{confraIEEE}, \cite{Schirmer}, for
%standard controllability cannot be adapted directly in this case.

\vs

According to classical  results of geometric control theory
\cite{JS} applied to quantum mechanics \cite{Butsam}, \cite{HT},
\cite{RSD}, the set of all available unitary transformations on the
total system $S+A$ is dense in the
 connected Lie group $e^{\cal L}$ corresponding to the Lie algebra ${\cal L}$
generated by the set \be{setkgkg} {\cal F}:=\{iH_S+iH_I+iH_A,\,
i {\bf 1} \otimes L_1, \ldots, i {\bf 1} \otimes L_m \},  \ee
 where $H_S,$ $H_I$ and $H_A$ were defined in (\ref{jh}) -
 (\ref{formaHC}), and it is equal to $e^{\cal L}$ if $e^{\cal L}$ is
 compact.\footnote{We shall not insist in the following on this
 distinction and in fact all the Lie groups we shall encounter are
 compact.} Therefore the Lie algebra ${\cal L}$ is of crucial
 importance in determining the structure of the reachable set for
 the  total system and the system $S$. In particular the set
 $\Lambda_S(\rho_S)$ defined in (\ref{reachi2}) can be written as
 \be{reachi2bis}
\Lambda_S(\rho_S)= \{\rho_S^{\prime}| \rho_A
\in \Lambda_A, X \in e^{\cal L}\}.
 \ee
The Lie algebra ${\cal L}$ is referred to as the {\it dynamical Lie
algebra} associated with the system. We shall also consider the {\it
control Lie algebra}, ${\cal B}$ which is generated by  $ \{i{\bf 1}
\otimes L_1, \ldots, i{\bf 1} \otimes L_m \}$; it turns out that ${\cal L}$ is
generated by ${\cal B}$ and $iH_S+iH_I+iH_A$.

\vs

%The  following theorem gives a general negative criterion in terms
%of the dynamical Lie algebra and an appropriate vector space for
%unitary
% indirect controllability given a set $\Lambda_A$ consisting of a
% single state $\rho_A$.

For a subspace ${\cal V}$ of ${\mathfrak {u}}(n_S n_A)$, let
${\rm Tr}_A({\cal V})$ denote the image of ${\cal V}$ under ${\rm Tr}_A$ in
${\mathfrak{u}}(n_S)$. Consider an initial state  for the total
system, $\rho_S \otimes \rho_A$, and the subspace of
${\mathfrak{u}}(n_S n_A)$, \be{observaspa} {\cal V}:=
\bigoplus_{k=0}^\infty {\rm ad}_{\cal L}^k \left( {\rm span}\{i \rho_S
\otimes \rho_A \}\right), \ee where ${\rm ad}_{\cal L}^0 {\cal P}={\cal
P}$ for any space ${\cal P}$ and, recursively, ${\rm ad}_{\cal L}^k {\cal
P}:=[{\cal L}, {\rm ad}_{\cal L}^{k-1} {\cal P} ]$.\footnote{The space
${\cal V}$ can be calculated recursively from a basis of ${\cal L}$,
and because of ${\mathfrak{u}}(n_Sn_A)$ being finite dimensional,
there exists a $\bar k$ such that ${\cal V}=\bigoplus_{k=0}^{\bar k}
{\rm ad}_{\cal L}^k \left( {{\rm span}}\{ i \rho_S \otimes \rho_A \}
\right)$.
%The space ${\cal V}$  is similar
%to the observability space defined in \cite{mikoobs} where the
%observable under consideration is replaced by $i \rho_S \otimes
%i\rho_A$.
The main (in fact, defining) property of ${\cal V}$ is that it is
the smallest subspace of ${\mathfrak{u}}(n_S n_A)$ which is
invariant under ${\cal L}$ and contains $i \rho_S \otimes \rho_A$.}

\bt{gennegat} Consider  the nontrivial case $\rho_S
\not=\frac{1}{n_S} \,{\bf 1}_{n_S}$. Assume that for all $X \in SU(n_S)$
there exists $U \in e^{\cal L}$ such that \be{tracc} {\rm Tr}_A (U \rho_S
\otimes \rho_A U^\dagger) = X\rho_S X^\dagger. \ee Then ${\rm Tr}_A({\cal
V})={\mathfrak{u}}(n_S)$.  \et
\bpr
Since ${\cal V}$ is invariant
under ${\cal L}$, it is also invariant under the action of $e^{\cal
L}$ by conjugation, i.e., $Q \in {\cal V} \Longrightarrow
UQU^\dagger \in {\cal V}$ for any $U \in e^{\cal L}$. Now, from
(\ref{tracc}) we have that, for every $X\in SU(n_S)$, $X i \rho_S
X^\dagger \in {\rm Tr}_A({\cal V}).$ This implies that \be{spanno}
{\rm span} \{X i\rho_S X^{\dagger} | X \in SU(n_S)\} \subseteq
{\rm Tr}_A({\cal V}). \ee Since $\rho_S \not=\frac{1}{n_S}{\bf
1}_{n_S}$, ${\rm span} \{X i\rho_S X^{\dagger} | X \in
SU(n_S)\}={\mathfrak{u}}(n_S)$, it follows ${\mathfrak{u}}(n_S) =
{\rm Tr}_A({\cal V})$.
\epr

Theorem \ref{gennegat} can be applied without calculating ${\cal V}$
when we recognize that $i\rho_S \otimes \rho_A$ belongs to a
subspace $\tilde {\cal V}$  invariant under commutation with ${\cal
L}$, which might in general include properly ${\cal V}$. If ${\rm Tr}_A
(\tilde {\cal V}) \not={\mathfrak{u}}(n_S)$, we can exclude the UIC
property. One case is when $i \rho_S \otimes \rho_A \in  {\cal L}
\oplus {\rm span}\{i {\bf 1}_{n_S} \otimes {\bf 1}_{n_A}\}$ and
${\rm Tr}_A ({\cal L} \oplus {\rm span}\{i {\bf 1}_{n_S} \otimes {\bf
1}_{n_A}\}) \not={\mathfrak{u}}(n_S)$. An application of this will
be used in Theorem \ref{UICequiv}. Another possibile case is when
$\tilde {\cal V}= {\cal L}^\perp \oplus i{\bf 1}_{n_S} \otimes {\bf
1}_{N_A}$, which is also invariant under ${\cal L}$. We have found
an example showing that the  converse of Theorem \ref{gennegat} is
not true in general and it is an interesting open problem to
understand how the condition of the theorem can be modified to have
a necessary and sufficient condition.

\section{THE DYNAMICAL LIE ALGEBRA FOR
THE INDIRECT CONTROL OF TWO QUBITS} \label{2qbit}

In this and the following section, we explore the case in which
target $S$ and accessor $A$ are both two-level systems. We start
with a study of the dynamical Lie algebra in all cases. We shall
treat separately the two cases in which the control Lie algebra
${\cal B}$ is the full ${\bf 1} \otimes {\mathfrak{su}}(2)$, and in
which it is one-dimensional. These are the only possible cases since
${\cal B}$ is always a Lie subalgebra of (a Lie algebra isomorphic
to) ${\mathfrak{su}}(2)$ and there  are no two-dimensional
subalgebras in ${\mathfrak{su}}(2)$. We shall also assume throughout
that the interaction Hamiltonian $H_I$ is non-zero since the case of
trivial interaction is clearly non controllable and consists of the
two systems evolving separately.
%In  the
%following section we shall see the  impacts of this study on the
%indirect controllability of the system.
%We shall see, in particular,
%that already in this case, there are examples where indirect
%controllability is possible in an appropriate sense without complete
%controllability.

\subsection{Algebra with Pauli matrices}

We recall for future reference the definition of the Pauli matrices
\be{tildesig} \tilde \sigma_x:= \pmatrix{0 & 1 \cr 1 & 0}, \, \,
\tilde \sigma_y:= \pmatrix{0 & i \cr -i & 0}, \, \, \tilde
\sigma_z:= \pmatrix{1 & 0 \cr 0 & -1}. \ee The related matrices
\be{Paulimat} \sigma_x := \frac{i}{2} \, \tilde \sigma_x, \quad
\sigma_y := \frac{i}{2} \, \tilde \sigma_y, \quad \sigma_z := \frac{i}{2} \, \tilde
\sigma_z, \ee span the standard two dimensional representation of
${\mathfrak{su}}(2)$ and satisfy the commutation
relations\footnote{The commutator $[B,C]$ is defined as
$[B,C]:=BC-CB$ and the anti-commutator $\{B,C\}$ is defined as
$\{B,C\}:=BC+CB$.}\be{commutrel} [\sigma_x,\sigma_y]=\sigma_z, \quad
[\sigma_y,\sigma_z]=\sigma_x, \quad [\sigma_z,\sigma_x]=\sigma_y,
\ee and the anti-commutation relations \be{anticommut} \{ \sigma_j,
\sigma_k \}= - \frac{1}{2} \, \delta_{jk} {\bf 1},\quad \{ \sigma_j, {\bf
1}\}=2\sigma_j  \qquad j,k=x,y,z. \ee Anti-commutation relations are
useful when using the formula \be{usefulformula} [A \otimes B, C
\otimes D]=\frac{1}{2} [A,C] \otimes \{B,D\} + \frac{1}{2} \{A,C\}
\otimes [B,D] \ee in calculations.

\subsection{Model for two interacting qubits}

%Assuming  ${\cal B}=su(2)$ (which includes the cases where there are
%two or three independent controls in (\ref{formaHC}))  we shall
%characterize the dynamical Lie algebra ${\cal L}$ in all cases, thus
%extending a result in \cite{Fu} which only gives a sufficient
%condition for complete controllability. We shall see that this
%condition is not necessary.
We specialize to the two qubits case the general model (\ref{jh}),
(\ref{HIgenform}), (\ref{formaHC}), and write ${\bf 1} = {\bf 1}_2$.
The Hamiltonian of the system $S$ has the form $H_S = B \otimes {\bf
1}$, with $iB \in {\mathfrak{su}}(2)$,  and we can make a change of
coordinates (cf. Remark \ref{Invariance}) to diagonalize $B$. This
does not modify the form of $H_I$ nor the dimension of  ${\cal B}$.
The Hamiltonian of the accessor system alone has the form $H_A:={\bf
1} \otimes C$ with $iC \in {\mathfrak{su}}(2)$. Therefore we have
\be{HSM} iH_S:=\omega_S \sigma_z \otimes {\bf 1},  \ee \be{HIM}
iH_I:=i\sigma_a \otimes \sigma_x+ i \sigma_b \otimes \sigma_y + i
\sigma_c \otimes \sigma_z, \ee \be{HAM} iH_A:= i {\bf 1} \otimes C,
\qquad i C \in {\mathfrak{su}}(2). \ee Here and in the following,
$\sigma_a$, with $ a\not= x,y,z$ indicates a matrix determined by a
vector $\vec a:=[a_x,a_y,a_z]^T \in \RR^3$ and given by \be{tty}
\sigma_a:=a_x \sigma_x + a_y \sigma_y + a_z \sigma_z, \ee that is a
general matrix in ${\mathfrak{su}}(2)$.\footnote{In this
correspondence between vectors in $\RR^3$ and elements in
${\mathfrak{su}}(2)$, we have $\sigma_x \leftrightarrow \vec i$,
$\sigma_y \leftrightarrow \vec j$, $\sigma_z \leftrightarrow \vec
k$. This correspondence between ${\mathfrak{su}}(2)$ and $\RR^3$ is,
in fact, a Lie algebra isomorphism, where in $\RR^3$ the Lie bracket
is replaced by the cross product.There is a correspondence also
between inner products in ${\mathfrak{su}}(2)$ and $\RR^3$ and
orthogonal vectors in $\RR^3$ correspond to orthogonal matrices in
${\mathfrak{su}}(2)$.} Define the matrix \be{matrixK}
K:=\pmatrix{\vec a^{\,\, T} \cr \vec b^{\, \, T} \cr \vec c^{\, \,
T}},  \ee and write it as \be{Kform} K:=\pmatrix{D & F}, \ee with
$D$ of dimension $3 \times 2$ and $F$ of dimension $3 \times
1$.\footnote{Notice that since we have assumed nontrivial
interaction at least one between $F$ and $D$ must be nonzero.}

In this notation the model is determined by the matrix $K$, the
parameter  $\omega_S$, the skew-Hermitian matrix $iC$ and the
control Lie algebra ${\cal B}$. We are interested in the dynamical
Lie algebra generated by the set \be{genLAM} {\cal S}:= \{\omega_S
\sigma_z \otimes {\bf 1} +i \sigma_a \otimes \sigma_x +i \sigma_b
\otimes \sigma_y +i \sigma_c \otimes \sigma_z+ i {\bf 1} \otimes C
\, , \, {\cal B} \}.  \ee In the following subsection we
characterize ${\cal L}$ in all cases when ${\cal B}$ is the full Lie
algebra ${\bf 1} \otimes {\mathfrak{su}}(2)$. Then,  we shall give a
necessary and sufficient controllability condition for the case
where ${\cal B}$ is $1$-dimensional and $\omega_S=0$, that is, the
system $S$ does not have dynamics by itself.

%and we are interested in the dynamical Lie algebra ${\cal L}$, which
%is generated by \be{DLM}
%\ee
%\ee
%First, observe that since ${\cal B}={\bf 1} \otimes su(2)$,
%$iH_A \in {\cal B}$ and the Lie algebra generated by
%$iH_S+iH_I+iH_A$ and ${\cal B}$ is the same as the one generated by
%$iH_S +iH_I$ and ${\cal B}$. Moreover $ Therefore, we can always,
%without loss of generality, consider the form for $iH_S+iH_I+iH_A$,
%\be{formwlg} iH_S+iH_I+iH_A= \omega_S \sigma_z \otimes {\bf 1} +
%i\sigma_a \otimes \sigma_x+ i \sigma_b \otimes \sigma_y + i \sigma_c
%\otimes \sigma_z.   \ee

\subsection{The dynamical Lie algebra in the case ${\cal B}={\mathfrak{su}}(2)$}

\bt{gendynLiealg} Assume $\dim {\cal B} =3$. Let $K$, $D$, $F$ and
$\omega_S$  defined as above. Then we have the following cases for
the dynamical Lie algebra ${\cal L}$ generated by ${\cal S}$ in
(\ref{genLAM}). If $\omega_S \not=0$, we have:

\vs

\noindent {\bf 1a)} If $D \not=0$ and $F\not=0$, the system is
completely controllable, that is, ${\cal L}={\mathfrak{su}}(4)$.

\vs

\noindent {\bf 1b}) If $D\not=0$ and $F=0$, then \be{calL1b} {\cal
L}= {\rm span} \{ \sigma_z \otimes {\bf 1}, {\bf 1} \otimes
{\mathfrak{su}}(2), i (\sigma_x, \sigma_y) \otimes
{\mathfrak{su}}(2) \}, \ee which is $10$-dimensional.

\vs

\noindent {\bf 1c}) If $D=0$ and $F \not=0$,  then \be{calL1c} {\cal
L}= {\rm span} \{ i\sigma_z \otimes {\mathfrak{su}}(2), \sigma_z
\otimes {\bf 1}, {\bf 1} \otimes {\mathfrak{su}}(2) \},  \ee which
is $7$-dimensional.

\vs

\noindent If $\omega_S=0$, we have:

\vs

\noindent {\bf 2a)} If ${\rm rank} (K) =1$, there exists $ \sigma
\in {\mathfrak{su}}(2)$ such that \be{calL2a} {\cal L}:=
{\rm span} \{ i \sigma \otimes {\mathfrak{su}}(2), {\bf 1}
\otimes {\mathfrak{su}}(2) \}. \ee

\vs

\noindent {\bf 2b}) If ${\rm rank} (K) =2$, ${\cal L}$ is
conjugate (and therefore isomorphic) to the Lie algebra ${\cal L}$
 of point 1b) above.

\vs

\noindent {\bf 2c}) If ${\rm rank} (K) =3$, the system $S+A$ is
completely controllable.

%\noindent {\bf 2c}) If $\det (K) \not=0$, then the system $S+A$ is
%completely controllable.

\et

The condition ${\rm rank} (K) =3$, or equivalently $\det (K)
\not=0$, is the complete controllability sufficient condition of
\cite{Fu} for both the cases $\omega_S=0$ and $\omega_S \not=0$. It
is clear that there are cases included in 1a) where this condition
is not verified and the system is nevertheless completely
controllable.

It is interesting to analyze the structure of the Lie algebra ${\cal
L}$ in the cases which are not completely controllable, in
particular, in cases 1b)
and 2b). This Lie algebra is isomorphic (and therefore conjugate,
see Lemma 4.2 in \cite{confraIEEE}) to ${\mathfrak{sp}}(2)$ as it can be shown
for example by using the test of Theorem 8 in \cite{confraIEEE}.
Therefore the total system $S+A$ is {\it pure state controllable},
which means that every transfer is possible between {\it pure
states} in ${\cal S} \otimes {\cal A}$. This implies that, if $\rho_S
\otimes \rho_A$ is a pure state, and the final desired state for
$\rho_S$ is also a pure state, this transfer will be possible in
the indirect control scheme. We shall see in Proposition \ref{pure9}
that a stronger property actually holds in this case, that is, Pure
UIC. We shall consider this case also in the Example
\ref{contoesempio}, in a setting of interest for applications. We now
give the proof of Theorem \ref{gendynLiealg}.

\bpr We notice that since ${\cal B}={\bf 1} \otimes
{\mathfrak{su}}(2)$ we can cancel ${\bf 1} \otimes C$ in (\ref{HAM})
with an element of ${\cal B}$. We also  notice that, in any case,
the dynamical Lie algebra ${\cal L}$ is the same as the Lie algebra
${\cal L}'$ generated by the set
\begin{eqnarray}\label{generaset}
{\cal S}' &:=& \{\omega_S \sigma_z \otimes {\bf 1}, i \sigma_a \otimes
{\mathfrak{su}}(2), \nonumber \\
&& \,\, i \sigma_b \otimes {\mathfrak{su}}(2), i
\sigma_c \otimes {\mathfrak{su}}(2), {\bf 1} \otimes
{\mathfrak{su}}(2)\}.
\end{eqnarray}
The fact that ${\cal L} \subseteq {\cal
L}'$ is obvious since the generators of ${\cal L}$ can be obtained
as linear combinations of the generators of ${\cal L}'$. To show the
converse inclusion we need to show that every generator of ${\cal
L}'$ listed in (\ref{generaset}) is in ${\cal L}$. To do this take
the Lie bracket of $\omega_S \sigma_z \otimes {\bf 1} + i\sigma_a
\otimes \sigma_x+i \sigma_b \otimes \sigma_y + i \sigma_c \otimes
\sigma_z \in {\cal L}$ with ${\bf 1} \otimes \sigma_z$. This gives
$-i \sigma_a \otimes \sigma_y + i \sigma_b \otimes \sigma_x$. Taking
the Lie bracket of this with ${\bf 1} \otimes \sigma_x$ gives $i
\sigma_a \otimes \sigma_z$ and with Lie brackets with ${\bf 1}
\otimes \sigma_{x,y}$ and linear combinations,  we obtain all
elements of the form $i\sigma_a \otimes \sigma$, with arbitrary
$\sigma \in {\mathfrak{su}}(2)$. Analogously we obtain elements in
$i\sigma_b \otimes {\mathfrak{su}}(2)$ and $i\sigma_c \otimes
{\mathfrak{su}}(2)$. We therefore analyze the Lie algebra generated
by the set in (\ref{generaset}) in all cases.

Assume first $\omega_S \not=0$. Assume  both $D$ and $F$ are
different from zero. Since $D \not=0$,  there exists one among
$\sigma_a,$ $ \sigma_b$ and $\sigma_c$ with  a nontrivial component
perpendicular to $\sigma_z$. Assume, without loss of generality,
that it is $\sigma_a$. We can perform a rotation (i.e., a unitary
similarity transformation) on the first qubit so that this component
is proportional to $\sigma_y$ without modifying $\sigma_z$, that is,
we can assume without loss of generality that $\sigma_a= \alpha
\sigma_y+ \beta \sigma_z$, with $\alpha \not=0$ which we can
therefore assume equal to $1$. Taking the Lie bracket of $i\sigma_a
\otimes \sigma $ for any $\sigma \in {\mathfrak{su}}(2)$ with
$\sigma_z \otimes {\bf 1}$, we obtain therefore $i\sigma_x \otimes
\sigma$. Taking the Lie bracket with $\sigma_z \otimes {\bf 1}$
again, we obtain $i \sigma_y \otimes \sigma$. This means that the
span of $i\sigma_x \otimes \sigma$ and $i \sigma_y \otimes \sigma$
is in ${\cal L}$. Now since $F \not=0$ there is one of the
$\sigma_{a,b,c}$ which has nonzero component along $\sigma_z$
(recall that our change of coordinates did not modify the components
along $\sigma_z$ since it amounts to a rotation in the $x-y$ plane).
Therefore we have every element of the form $i\sigma_1 \otimes
\sigma_2$, $\sigma_1 \in {\mathfrak{su}}(2)$ and $\sigma_2 \in
{\mathfrak{su}}(2)$. The remaining terms of the standard basis of
${\mathfrak{su}}(4)$ can be obtained by Lie brackets of the type
$[i\sigma_1 \otimes \sigma_z, i\sigma_2 \otimes \sigma_z]$, for
appropriate $\sigma_1$ and $\sigma_2$ in ${\mathfrak{su}}(2)$. In
the case $D \not=0$ but $F=0$, the argument is the same but it stops
at the point where we have generated all elements in the Lie algebra
described in (\ref{calL1b}). If $D=0$ but $F\not=0$, there is at
least one element among $\sigma_a$, $\sigma_b$ and $\sigma_c$ which
has nonzero component along $\sigma_z$ and this is the only possible
nonzero component. Therefore we have already all the elements listed
in (\ref{calL1c}) which span a proper Lie subalgebra of
${\mathfrak{su}}(4)$.

Consider now the case where $\omega_S=0$, i.e., there is no natural
dynamics of the target system $S$. If ${\rm rank} ([D \quad
F])=1$ then there exists a $\sigma \in {\mathfrak{su}}(2)$ such that
$\sigma_{a,b,c}=k_{a,b,c} \sigma$, for coefficients $k_{a,b,c}$.
Therefore we obtain the Lie algebra described in (\ref{calL2a}).  If
the rank is $2$ there are two among $\sigma_a$, $\sigma_b$ and
$\sigma_c$ which are linearly independent while the third one is a
linear combination of these two and therefore the corresponding
matrices $M \otimes   \sigma$, with $M$ free,  do not contribute to
the Lie algebra. Assume, without loss of generality that the
linearly independent matrices are $\sigma_a$ and $\sigma_b$. By a
rotation again, up to a nonzero unimportant proportionality factor,
we can assume that $\sigma_a=\sigma_x$ and $\sigma_b=\sigma_y+\alpha
\sigma_x$. Therefore $i(\sigma_x,\sigma_y) \otimes
{\mathfrak{su}}(2)$ is included in the Lie algebra ${\cal L}$. By
taking the Lie bracket $[i\sigma_x \otimes \sigma_z, i \sigma_y
\otimes \sigma_z]$ one obtains the remaining basis element $\sigma_z
\otimes {\bf 1}$ in (\ref{calL1b}). In the case 2c) (rank 3) one
adds to the previously generated elements another set of elements
$i\sigma_c \otimes \sigma$ with arbitrary $\sigma \in
{\mathfrak{su}}(2)$ and notice that $\sigma_c$ must have, in the
chosen coordinates, non zero component along $\sigma_z$. Therefore
one obtains as above the remaining elements in the standard basis of
${\mathfrak{su}}(4)$. Alternatively, and more quickly, one can use
the fact pointed out after the statement of the theorem, that the
Lie algebra (\ref{calL1b}) is conjugate to ${\mathfrak{sp}}(2)$ and every such
Lie algebra is known to be a maximal subalgebra \cite{Dynkin} in
${\mathfrak{su}}(4)$, that is, the addition of every nonzero element
outside the Lie algebra but inside ${\mathfrak{su}}(4)$, causes the
generation of the whole Lie algebra ${\mathfrak{su}}(4)$. \epr

\subsection{{Complete controllability when  ${\cal B}$ is 1-dimensional and $\omega_S=0$}}

In the case where the control Lie algebra ${\cal B}$ is
one-dimensional, that is, there is only one independent  control,
the analysis becomes more complicated. Among the other issues,  one
has to consider that the independent dynamics of the accessor system
(i.e., the Hamiltonian $H_A$ in (\ref{jh})) cannot be canceled in
general by the action of the control and play a significant role in
determining the dynamics of the total system. Nevertheless, we can
use the result of Theorem \ref{gendynLiealg} jointly with the fact
that, with the same parameters for the system, the dynamical Lie
algebra in the case of ${\cal B}$ $1$-dimensional is always a
subalgebra (not necessarily proper) of the corresponding dynamical
Lie algebra for the case of ${\cal B}$ $3$-dimensional. We restrict
ourselves to the case where $\omega_S=0$.

\bt{OMS0} Assume $\omega_S=0$ and $\dim {\cal B}=1$. Then the system
$S+A$ is completely controllable if and only if the following two
conditions are verified:

C1) $\det(K) \not=0$;

C2) The components of ${\rm Tr}_A(iH_I H_C)$ and ${\rm Tr}_S(H_A)$ perpendicular
to ${\rm Tr}_S(H_C)$ are not both zero.

\et

Notice that condition C1 is the same as the condition 2c in the
$\dim {\cal B}=3$ case. To that, we have to add the generically
satisfied condition C2 to have a necessary and sufficient condition
in the case of ${\cal B}$ $1-$dimensional.

\bpr  It is useful  to make a change of coordinates on the system
$A$, so as to make ${\cal B}={\rm span} \{ {\bf 1} \otimes
\sigma_z \}$. In these coordinates, the part of $iC$ in
(\ref{genLAM}) which is parallel to $\sigma_z$ can be neglected
because it is already contained in ${\cal B}$. Therefore, we can
consider only the part perpendicular to $\sigma_z$ which can be
taken equal to $\omega_A\sigma_y$, for a certain real parameter
$\omega_A$. In these coordinates, we need to redefine $\sigma_a$,
$\sigma_b$ and $\sigma_c$ and therefore $K$. Moreover we can make a
change of coordinates on the system $S$ to make $\sigma_b=\beta
\sigma_y$, $\sigma_a=(\alpha \sigma_x+ \gamma \sigma_y)$ and
$\sigma_c=x \sigma_x+y\sigma_y+z\sigma_z$, for real coefficients
$\alpha,\beta,\gamma$ and $x,y,z$.

\noindent Conditions C1 and C2, in these coordinates, become

C1') \be{detco1} \det \pmatrix{\alpha & \gamma & 0 \cr 0 & \beta & 0
\cr x & y & z} \not=0.  \ee

C2') \be{detco2} \omega_A^2+x^2+y^2 \not=0.  \ee The condition
(\ref{detco1}) is necessary because it is needed in the
 ${\cal B}$ 3-dimensional case.
 If condition (\ref{detco2}) is not satisfied,
the set ${\cal S}$ in (\ref{genLAM}) can be taken, in these
coordinates,  \be{genLAMspec} {\cal S}:=\{ i\alpha \sigma_x \otimes
\sigma_x + i \gamma \sigma_y \otimes \sigma_x +i \beta \sigma_y
\otimes \sigma_y +i z \sigma_z \otimes \sigma_z, {\bf 1} \otimes
\sigma_z \},  \ee and both elements in this set are in the Lie
subalgebra
\begin{eqnarray}\label{subalg}
{\cal L}' &:=& {\rm span}\{ {\bf 1} \otimes
\sigma_z, \sigma_z \otimes {\bf 1}, i \sigma_y \otimes \sigma_x, i
\sigma_x \otimes \sigma_y, \nonumber \\
&& \qquad \quad i \sigma_y \otimes \sigma_y, i \sigma_x
\otimes \sigma_x, i \sigma_z \otimes \sigma_z\}.
\end{eqnarray}

We now prove the sufficiency of the conditions C1 and C2 for
complete controllability. We examine the Lie algebra generated by
the set ${\cal S}$ which can be written in the given coordinates
\be{setge1} {\cal S}:=\{ i \sigma_a \otimes \sigma_x +i \sigma_b
\otimes \sigma_y + i \sigma_c \otimes \sigma_z+ \omega_A {\bf 1}
\otimes \sigma_y, {\bf 1} \otimes {\sigma_z}\}.  \ee We notice
that the Lie algebra generated by ${\cal S}$ in (\ref{setge1}) is
the same as the Lie algebra generated by the set with three elements
${\cal S}'$ defined as \be{setge2} {\cal S}':=\{ i \sigma_c \otimes
\sigma_z, i \sigma_a \otimes \sigma_x +i \sigma_b \otimes
\sigma_y + \omega_A {\bf 1} \otimes \sigma_y, {\bf 1} \otimes
\sigma_z \}.  \ee This can be easily seen because, if we take the
Lie bracket of the first element of  ${\cal S}$ in (\ref{setge1})
with the second element (${\bf 1} \otimes \sigma_z$) two times, we
obtain the second element in ${\cal S}'$ in (\ref{setge2}), and by
subtracting this from the first element of ${\cal S}$, we obtain the
first element in ${\cal S}'$. We first characterize the Lie algebra
${\cal L}''$ generated by the second and third element in ${\cal
S}'$ in (\ref{setge2}), i.e., the Lie algebra generated by
\be{setge3} {\cal S}{''}:=\{ i \sigma_a \otimes \sigma_x +i \sigma_b
\otimes \sigma_y+ \omega_A {\bf 1} \otimes \sigma_y, {\bf
1}\otimes \sigma_z \}. \ee Then we examine what happens when we
include the first element of ${\cal S}'$ and  see that, under the
given assumptions, we generate the whole Lie algebra
${\mathfrak{su}}(4)$.

\noindent In the given coordinates, we have
\begin{eqnarray}\label{esse1bis}
{\cal S}' &:=& \{ i(x \sigma_x +y \sigma_y +z \sigma_z) \otimes \sigma_z, {\bf 1} \otimes \sigma_z, \\
&& \,\, i (\alpha \sigma_x + \gamma \sigma_y) \otimes \sigma_x +i \beta \sigma_y
\otimes \sigma_y + \omega_A {\bf 1} \otimes \sigma_y \}, \nonumber
\end{eqnarray}
and ${\cal S}''$ is given by the last two terms listed in (\ref{esse1bis}).

\noindent Define \be{cond2b}
 k:= \alpha^2+4\omega_A^2,
 \ee
which is always non-vanishing since $\alpha \not=0$ from
(\ref{detco1}).

\noindent Define the following matrices:
 \be{gammax}
\Gamma_x^{\pm}:=\left( i\sigma_y \otimes \sigma_x \pm
\frac{1}{\sqrt{k}}(\alpha i \sigma_x \otimes \sigma_y - \omega_A
{\bf 1} \otimes \sigma_x )\right),
 \ee
\be{gammay} \Gamma_y^{\pm}:=- \frac{1}{2} \left( {\bf 1} \otimes
\sigma_z \pm \frac{1}{\sqrt{k}}(\alpha \sigma_z \otimes {\bf 1} -4
\omega_A i\sigma_y \otimes \sigma_z) \right),  \ee

\be{gammaz} \Gamma_z^{\pm}:= \left(  i\sigma_y \otimes \sigma_y \mp
\frac{1}{\sqrt{k}} (\alpha i \sigma_x \otimes \sigma_x + \omega_A
{\bf 1} \otimes \sigma_y ) \right).  \ee It is straightforward  to
verify that \be{commurel1} [\Gamma_x^+, \Gamma_y^+]=\Gamma_z^+,
\quad [\Gamma_y^+, \Gamma_z^+]= \Gamma^+_x, \quad  [\Gamma_z^+,
\Gamma_x^+]= \Gamma^+_y,  \ee and \be{commurel2} [\Gamma_x^-,
\Gamma_y^-]=\Gamma_z^-, \quad [\Gamma_y^-, \Gamma_z^-]= \Gamma^-_x,
\quad  [\Gamma_z^-, \Gamma_x^-]= \Gamma^-_y.  \ee Therefore, the Lie
algebras ${\cal L}^+:= {\rm span} \{ \Gamma_x^+,  \Gamma_y^+,
\Gamma_z^+ \}$ and ${\cal L}^-:= {\rm span} \{ \Gamma_x^-,
\Gamma_y^-,  \Gamma_z^- \}$, are both isomorphic to
${\mathfrak{su}}(2)$ (cf. (\ref{commutrel})), and one can verify
that \be{commu0} [{\cal L}^+, {\cal L}^-]=0, \ee so that the sum
${\cal L}^{+} \oplus {\cal L}^-$ is, in fact, a direct sum of Lie
algebras.
%We want to show
%that in any case the Lie algebra ${\cal L}"$ is a subalgebra of a
%Lie algebra os the form ${\cal L}^+ \oplus {\cal L}^-$.
%We claim that ${\cal L}''$ is a subalgebra of ${\cal L}^+ \oplus
%{\cal L}^-$ with equality verified if and only if conditions
%(\ref{cond1}), (\ref{cond2}) and (\ref{cond3})  are verified.
Consider the generators of ${\cal L}''$. We have \be{calcu1} L_1=
{\bf 1} \otimes \sigma_z= - (\Gamma_y^+ + \Gamma_y^-), \ee and
\begin{eqnarray}\label{calcu2}
L_2 &=& i \alpha \sigma_x \otimes \sigma_x + i \gamma
\sigma_y \otimes \sigma_x + i \beta \sigma_y \otimes \sigma_y +
\omega_A {\bf 1} \otimes \sigma_y \nonumber \\
&=& \frac{1}{2} \Bigl[\gamma (\Gamma_x^+ + \Gamma_x^-) + \sqrt{k} (\Gamma_z^- -
\Gamma_z^+) + \beta (\Gamma_z^+ + \Gamma_z^-)\Bigr]. \nonumber \\
\end{eqnarray}
From these expressions it follows that ${\cal L}''$ is always a
subalgebra of  ${\cal L}^+ \oplus {\cal L}^-$. Using
(\ref{commurel1}), (\ref{commurel2}), (\ref{commu0}), we calculate
\begin{eqnarray}\label{conto1}
[[L_1,L_2],L_2] &=& \frac{1}{4} \Bigl[\left(\gamma^2 + (\beta -\sqrt{k})^2
\right) \Gamma_y^+ + \nonumber \\
&& \quad + \left( \gamma^2+(\beta + \sqrt{k})^2 \right) \Gamma_y^- \Bigr],
\end{eqnarray}
and this, along with the fact that $\beta$ and $k$
are both different from zero, and the expression of $L_1$ in
(\ref{calcu1}), shows that both $\Gamma_y^+$ and $\Gamma_y^-$ belong
to ${\cal L}''$. Using this fact, it is easy to verify, using
(\ref{commurel1}) and (\ref{commurel2}), that there are only three
possibilities:

\begin{enumerate}

\item $\gamma=0$ and $\beta=\sqrt{k}$
and ${\cal L}''$ is given by \be{calL2p} {\cal L}''={\rm span}\{
\Gamma_x^-, \Gamma_y^-,\Gamma_z^-, \Gamma_y^+\} \ee

\item
$\gamma=0$ and $\beta=-\sqrt{k}$  and ${\cal L}''$ is given by
\be{calL2s} {\cal L}''={\rm span}\{ \Gamma_x^+, \Gamma_y^+,
\Gamma_z^+, \Gamma_y^-\} \ee

\item At least one between $\gamma$ and $|\beta|-\sqrt{k}$ is
different from zero and ${\cal L}''$ is equal to ${\cal L}^+ \oplus
{\cal L}^-$.

\end{enumerate}

The task is to examine, under the given assumptions, what happens
when we add to these Lie algebras the matrix $ix\sigma_x \otimes
\sigma_z + iy \sigma_y \otimes \sigma_z + iz \sigma_z \otimes
\sigma_z:=i\sigma_c \otimes \sigma_z$. Let us assume without loss of
generality that $\sigma_c$ is scaled so that
$\{\sigma_c,\sigma_c\}=-\frac{1}{2} \, {\bf 1}$ (cf.
(\ref{anticommut})). The case (\ref{calL2p}) is considered in the
Appendix. The case (\ref{calL2s}) is similar.

\vs

\noindent In the case where ${\cal L}''={\cal L}^+ \oplus {\cal L}^-$, we
calculate, defining $T:=i x \sigma_x \otimes \sigma_z +i y \sigma_y
\otimes \sigma_z + i z \sigma_z \otimes \sigma_z$, \be{calcolo5}
\left[ (\Gamma_z^++\Gamma_z^-), T \right]=\frac{y}{4} \, {\bf 1}
\otimes \sigma_x.   \ee If $y \not=0$, the result follows as before
by comparison with the case of ${\cal B}$ $3$-dimensional. If $y=0$
at least one between $x$ and $\omega_A$ is different from zero. We
calculate
\begin{eqnarray}\label{contori}
&& \frac{1}{\sqrt{k}} \left[ 8 \omega_A
[\Gamma_z^+,T] + \alpha x (\Gamma^+_z - \Gamma_z^-), T \right] = \nonumber \\
&& \qquad \qquad = \left( \frac{x^2}{4} + \frac{\omega_A^2}{k} z^2 \right) {\bf 1} \otimes \sigma_y,
\end{eqnarray}
and the result follows as before by comparison with
the case of ${\cal B}$ 3-dimensional. \epr

\section{INDIRECT CONTROLLABILITY FOR THE TWO QUBIT SYSTEM}
\label{indcon}

The following theorem says that  strong unitary indirect
controllability is equivalent to complete controllability for the
system of two qubits. Nonetheless, we shall see that weaker notions
of indirect controllability are still possible without complete
controllability.

\bt{UICequiv} A system of two qubits with ${\cal B}={\bf 1} \otimes
{\mathfrak{su}}(2)$ is strong UIC if and only if it is CC. \et

\bpr We have already noticed in Section \ref{general}  how complete
controllability implies strong UIC. Assume now that complete
controllability is not verified. The dynamical Lie algebra
%whether we are in the case ${\cal B}$ $1$-dimensional or ${\cal B}$
%3-dimensional
is (modulo similarity transformations acting separately on the two
qubits) one of the Lie algebras (\ref{calL1b}), (\ref{calL1c}) or
(\ref{calL2a}). Therefore, it is sufficient to consider these cases.
In the case (\ref{calL1c}), consider for $S$ an initial density
matrix of the form $\rho_S={\bf 1}+ \kappa i \sigma_z$. Every
element in ${\cal L}$ has the form \be{formaAA} A={\bf 1} \otimes
{\sigma_1} +a \sigma_z \otimes {\bf 1} + i \sigma_z \otimes
\sigma_2,  \ee with $a$ a real number, and $\sigma_1$ and $\sigma_2$
elements of ${\mathfrak{su}}(2)$. Since $e^{\cal L}$ is a compact
Lie group, every element $X$ can be written as the exponential of a
matrix $A \in {\cal L}$. Therefore we have \be{Lieformula} X \rho_S
\otimes \rho_A X^\dagger = e^{A} \rho_S \otimes \rho_A e^{-A}
= \sum_{j=0}^\infty  \frac{1}{j!} \, {\rm ad}_A^j \rho_S \otimes \rho_A. \ee
Now, by induction on $j$, given the form of $\rho_S$ as a diagonal
matrix, it is easily seen that ${\rm ad}^j_A \rho_S \otimes \rho_A$ is
always the sum of a finite number of tensor products where the first
factor is a diagonal element. Taking the partial trace with respect
to $A$ we still obtain a diagonal matrix. Therefore, from diagonal
density matrices we can never reach matrices which are not diagonal.
\footnote{Notice that this argument also holds if we consider
specific sets $\Lambda_A$ for $\rho_A$. Therefore UIC given
$\Lambda_A$ is never verified, no matter what $\Lambda_A$ is.}
% As an
%alternative proof, we could have applied Theorem \ref{gennegat},
%whose proof is based on a similar argument, leaving $\rho_A$ as a
%free parameter.}

In the case of ${\cal L}$ as in (\ref{calL2a}), the proof is the
same. In fact, the Lie algebra (\ref{calL2a}) becomes  a subalgebra
of (\ref{calL1c}) after  a change of coordinates on $S$ which
diagonalizes $\sigma$.

In the case of ${\cal L}$ as in (\ref{calL1b}), consider
$\rho_S=\frac{1}{2} ({\bf 1} + k i \sigma_z)$ for some real $k$.
With this choice,  $i\rho_S \otimes \rho_A$ belongs to ${\tilde {\cal
L}}= ({\rm span}(i{\bf 1} \otimes {\bf 1})) \oplus {\cal L} $
which is invariant under ${\cal L}$. The result follows by applying
Theorem \ref{gennegat}, as ${\rm Tr}_A ( \tilde {\cal L}) \not=
{\mathfrak{u}}(n_S)$ (see the discussion after the
theorem).\footnote{In the case (\ref{calL1b}), ${\cal L}$ is
conjugated to the symplectic Lie algebra ${\mathfrak{sp}}(2)$ and therefore any
transfer is possible between pure states. If we want to find an
example to show that strong UIC is not verified we have to use
$\rho_S \otimes \rho_A$ non-pure.} \epr

\vspace{0.25cm}

The proof of the Theorem in the cases (\ref{calL1c}), (\ref{calL2a})
works to show that UIC given a set $\Lambda_A$ is not verified, even
even if we fix  the set $\Lambda_A$ a priori. It is interesting to
investigate if this is the case also for (\ref{calL1b}). In the next
proposition we will show that if $\Lambda_A$ is chosen as the set of
all pure states, then the UIC property given $\Lambda_A$ is verified
in the  case (\ref{calL1b}). We first present an example in order to
physically motivate the study of the Lie algebra (\ref{calL1b}).
This example also illustrates the use of Lie group decomposition
techniques in a direct analysis of the reachable set
(\ref{reachi2bis}).

\bex{contoesempio} Consider the system of two qubits $S$ and $A$
interacting via an Ising interaction and with full control on the
system $A$. It is described by \be{HamilS} iH_S:=\sigma_z \otimes
{\bf 1}, \ee \be{HamilC} iH_C:= u_x(t) {\bf 1}\otimes \sigma_x +
u_y(t) {\bf 1} \otimes \sigma_y , \ee \be{HamilI} H_I:= \sigma_y
\otimes \sigma_y.  \ee An application of Theorem \ref{gendynLiealg}
(or a direct calculation) shows that the dynamical Lie algebra
${\cal L}$ is given by (\ref{calL1b}).\footnote{When analyzing a
quantum mechanical system whose dynamical Lie algebra ${\cal L}$ is
not the full ${\mathfrak{su}}(n)$ one approach is to first decompose
it according to the Levi decomposition \cite{deGraaf} into simple
and Abelian Lie subalgebras \cite{mikodeco}.  In our  case ${\cal
L}$ is isomorphic to ${\mathfrak{sp}}(2)$ and therefore it is simple. Levi
decomposition
 only  gives  a simple component and no Abelian subalgebras.} To
parametrize the elements of the corresponding Lie group $e^{\cal
L}$, we use the Cartan decomposition \cite{Helgason} of ${\cal L}$
\be{Cartan16} {\cal L}={\cal K} \oplus {\cal P},  \ee with
\begin{eqnarray}\label{calKP}
{\cal K} &:=& {\rm span} \{\sigma_z \otimes {\bf 1}, 1 \otimes
\sigma_z, i\sigma_x \otimes \sigma_z, i \sigma_y \otimes \sigma_z
\}, \nonumber \\
{\cal P} &:=& {\rm span}\{ {\bf 1} \otimes
\sigma_x, {\bf 1} \otimes \sigma_y, i \sigma_x \otimes \sigma_x, i \sigma_x \otimes \sigma_y, \nonumber \\
&& \qquad \quad i \sigma_y \otimes \sigma_x, i \sigma_y
\otimes \sigma_y\},
\end{eqnarray}
satisfying \be{basiccommurel} [{\cal K},
{\cal K}]\subseteq {\cal K}, \quad [{\cal P}, {\cal P}] \subseteq
{\cal K}, \quad [{\cal P}, {\cal K}] \subseteq {\cal P}. \ee Every
element $X$ in $e^{\cal L}$ can be written as $X=K_1 \tilde A K_2$
where $K_1$ and $K_2$ are in the Lie group $e^{\cal K}$ and $\tilde
A$ is in the Lie group $e^{\tilde {\cal A}}$ where ${\tilde {\cal
A}}$ is a maximal Abelian subalgebra of ${\cal L}$ included in
${\cal P}$. Since ${\cal K}$ is isomorphic to ${\mathfrak{u}}(2)$,
we can use an Euler decomposition for $K_1$ and $K_2$ and write
\be{K1} K_1= e^{t_1 \sigma_z \otimes {\bf 1}} e^{t_2 {\bf 1} \otimes
{\sigma_z}} e^{i t_3 \sigma_x \otimes \sigma_z } e^{t_4 \sigma_z
\otimes {\bf 1}}, \ee \be{K2} K_2= e^{s_1 \sigma_z \otimes {\bf 1}}
e^{i s_2 \sigma_x \otimes \sigma_z } e^{s_3 \sigma_z \otimes {\bf
1}} e^{s_4 {\bf 1} \otimes {\sigma_z}},   \ee for real coefficients
$t_1,\ldots,t_4,s_1,\ldots,s_4$.  For the Abelian subalgebra $\tilde
{\cal A}$, we can choose \be{Abe} \tilde {\cal A}:={\rm span} \{
{\bf 1} \otimes \sigma_x, i \sigma_x \otimes \sigma_x \},  \ee so
that \be{Atil} \tilde A= e^{a_1 {\bf 1} \otimes \sigma_x }e^{ i a_2
\sigma_x \otimes \sigma_x },  \ee for real coefficients $a_1$ and
$a_2$. Notice that we have used the freedom in ordering  the factors
in a different way in (\ref{K1}) and (\ref{K2}), so as to minimize
the number of parameters which are relevant for the indirect
controllability problem. The first factor in (\ref{K1}) can be
pulled out of the partial trace (cf. (\ref{reachi2bis}) and
(\ref{mainpropertyPT})) while the second factor is a unitary local
operation on $A$ which does not affect the partial trace and can be
neglected. The last two terms in (\ref{K2}) can be included in the
initial state $\rho_S \otimes \rho_A$.
%Notice
%also that we can assume without loss of generality that $\rho_A$ is
%diagonal because all unitary operations are admissible on
%$A$.\footnote{Alternatively and equivalently we could have used a
%different Cartan decomposition where $\sigma_z$ is replaced by a
%matrix commuting with $\rho_A$}
In conclusion, we have to consider a transformation $\rho_S \otimes
\rho_A \rightarrow Y \rho_S \otimes \rho_A Y^\dagger$ with the
matrix $Y$ of the form
 \be{whyY} Y= e^{i t_3 \sigma_x \otimes \sigma_z } e^{t_4 \sigma_z \otimes
{\bf 1}} e^{a_1 {\bf 1} \otimes \sigma_x }e^{ i a_2 \sigma_x \otimes
\sigma_x } e^{s_1 \sigma_z \otimes {\bf 1}} e^{i s_2 \sigma_x
\otimes \sigma_z }. \ee The reachable set starting from $\rho_S
\otimes \rho_A$ will be
\begin{eqnarray}\label{reachi2tris}
\Lambda_S(\rho_S) &=& \left\{
e^{t_1 \sigma_z } ({\rm Tr}_A \Omega) \, e^{t_1 \sigma_z }| \, t_1, s_3, s_4 \in \RR \right\}, \\
\Omega &:=& \left( Y e^{s_3 \sigma_z } \rho_S e^{-s_3
\sigma_z } \otimes e^{s_4 \sigma_z } \rho_A e^{-s_4 \sigma_z }
Y^\dagger \right), \nonumber
\end{eqnarray}
with $Y$ varying as in (\ref{whyY}). Each term of $Y$
in (\ref{whyY}) can be expressed in terms of a basis of
$i{\mathfrak{u}}(4)$. We have for example \be{esa} e^{i t_3 \sigma_x
\otimes \sigma_z } = \cos{\alpha_1} {\bf 1} \otimes {\bf 1} +i \sin{\alpha_1}
\tilde \sigma_x \otimes \tilde \sigma_z, \quad
\alpha_1:=-\frac{t_3}{4}, \ee and the other expressions are
collected in the Appendix, see (\ref{add1}).
Combining all the expressions for the factors in (\ref{whyY}), we
obtain for $Y$ a formula of the type \be{hjl} Y:=C_0 \otimes {\bf 1}
+ C_x \otimes \tilde \sigma_x + C_y \otimes \tilde \sigma_y +C_z
\otimes \tilde \sigma_z,  \ee for matrices $C_0$, $C_x$, $C_y$, and
$C_z$, functions of $6$ real arbitrary parameters $\alpha_1, \ldots,
\alpha_6$. We have calculated the expressions of these matrices and
the results are reported in the Appendix.

\begin{figure}[t]
\centering
%\begin{center} % Requires \usepackage{graphicx}
  \includegraphics[width=8cm]{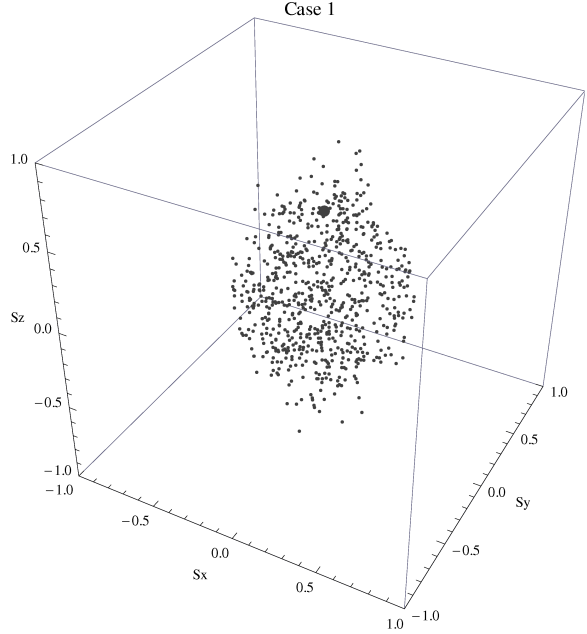} \\
 \caption{\footnotesize Simulation of the reachable set as described in the
main text, with $s_x = 0$, $s_z = 1/2$ and $a_z = 1$. The simulation
consists of $729$ points randomly distributed; the big spot marks
the initial state. The state of the accessor system is pure and it
is possible to reach points that are outside of the sphere
corresponding containing the initial state. This means that it is
possible to make the state more pure with the scheme studied here.
}\label{fig1}
%\end{center}
\end{figure}

\begin{figure}[t]
\centering
%\begin{center} % Requires \usepackage{graphicx}
  \includegraphics[width=8cm]{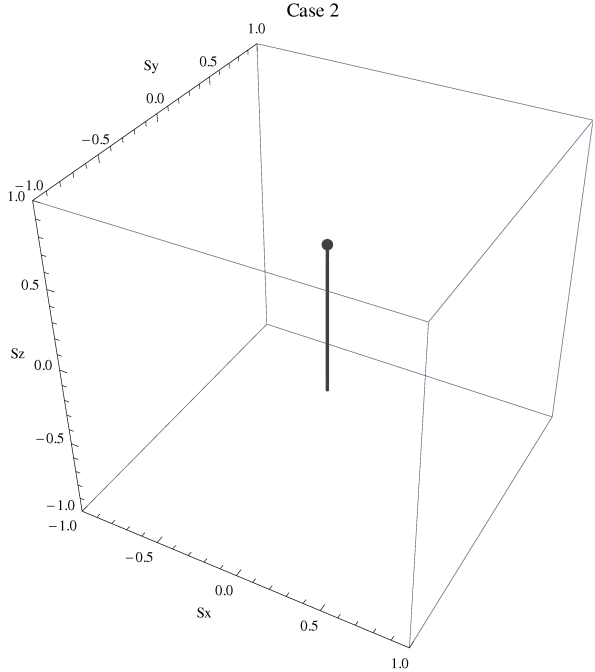} \\
 \caption{\footnotesize Simulation of the reachable set as described in the
main text, with $s_x = 0$, $s_z = 1/2$ and $a_z = 0$. The simulation
consists of $729$ points randomly distributed; the big spot marks
the initial state. In this case the initial state is in a space
invariant under the dynamical Lie algebra (modulo the identity it is
contained in ${\cal L}$) and remains in that space. This is
reflected in the picture where only a line parallel to the $z$ axis
is filled.}\label{fig2}
%\end{center}
\end{figure}

\begin{figure}[t]
\centering
%\begin{center} % Requires \usepackage{graphicx}
  \includegraphics[width=8cm]{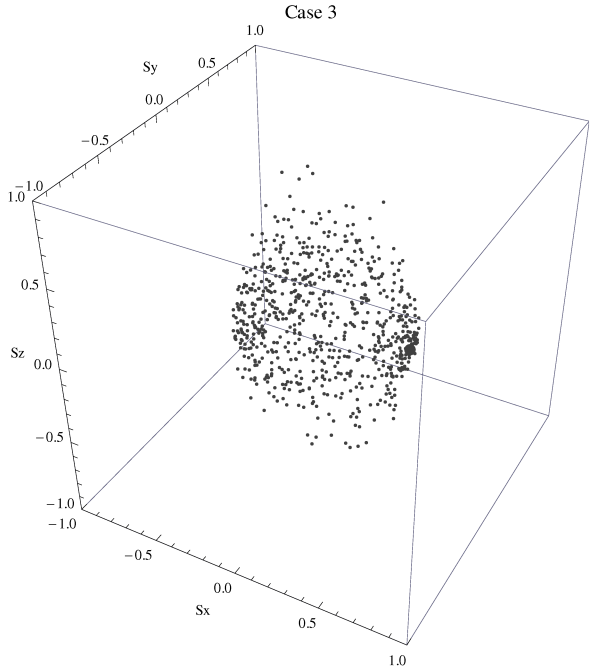} \\
 \caption{\footnotesize Simulation of the reachable set as described in the
main text, with $s_x = 1/2$, $s_z = 0$ and $a_z = 1$. The simulation
consists of $729$ points randomly distributed; the big spot marks
the initial state.}\label{fig3}
%\end{center}
\end{figure}

\begin{figure}[t]
\centering
%\begin{center} % Requires \usepackage{graphicx}
  \includegraphics[width=8cm]{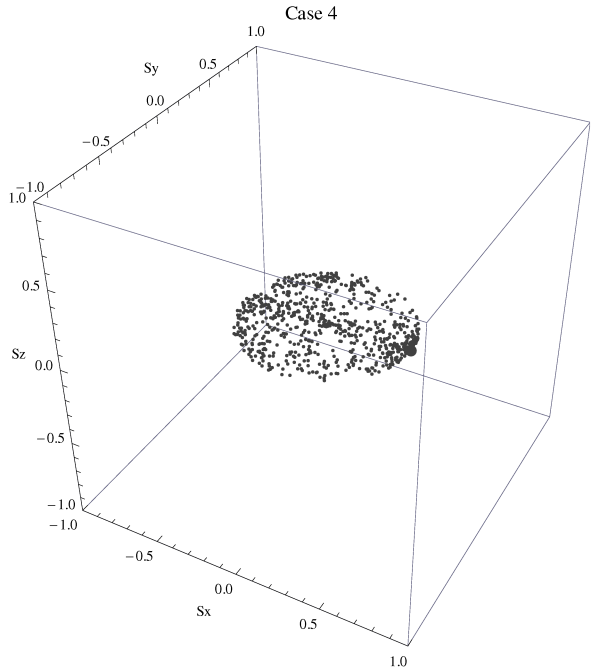} \\
 \caption{\footnotesize Simulation of the reachable set as described in the
main text, with $s_x = 1/2$, $s_z = 0$ and $a_z = 0$. The simulation
consists of $729$ points randomly distributed; the big spot marks
the initial state. The situation is similar to the one in Figure
\ref{fig2} since now the initial state is, modulo the identity, in
$i{\cal L}^\perp$ which is also invariant under ${\cal L}$. This
explain why all reachable points are in the $x-y$ plane.
}\label{fig4}
%\end{center}
\end{figure}

Formulas such as (\ref{reachi2tris}), (\ref{hjl}) allow in principle
to calculate an expression of the reachable set for every initial
state but in practice are typically very complicated and are useful
only with the help of numerical simulations. We have plotted the
reachable set starting from a state \be{inistat6} \rho_S \otimes
\rho_A=\left(\frac{1}{2}({\bf 1} + s_x \tilde \sigma_x + s_z \tilde
\sigma_z)\right) \otimes \left(\frac{1}{2}({\bf 1} + a_z \tilde
\sigma_z)\right) \ee in several situations, see Fig.
\ref{fig1}-\ref{fig4}. This illustrates the variety of cases that
can arise depending on the underlying algebraic conditions.

%They describe the reachable set in full generality. Much less is
%needed in the proof of the following proposition.
\eex

\bp{pure9} The system of two qubits with Lie algebra ${\cal L}$ in
(\ref{calL1b}) is pure UIC. \ep

\bpr Inspection of the basis of ${\cal L}$ in (\ref{calL1b}) tells
us that the Lie group $e^{\cal L}$ contains elements of the form
\be{formaelementi} Z \otimes {\bf 1}, \quad {\bf 1} \otimes P,\quad
e^{i \sigma_x \otimes \sigma_z t},  \ee where $Z$ is an arbitrary
element of the form $e^{\sigma_z \alpha}$ with $\alpha$  real, $P$
is any arbitrary element in $SU(2)$ and $t$ an arbitrary real
number. We start with an initial state for $\rho_S \otimes \rho_A$
with $\rho_A$ pure. By applying a transformation $ {\bf 1} \otimes P
$, we can always assume that \be{assunzionesura} \rho_A= \pmatrix{ 1
& 0 \cr 0 & 0}:=E_1 \ee Consider now a general matrix $\rho_S'
\otimes E_1$ and calculate
\begin{eqnarray}\label{calcolo3}
&& e^{i t \sigma_x \otimes \sigma_z} \rho_S' \otimes E_1  e^{-i t \sigma_x \otimes \sigma_z} = \nonumber \\
%\cos^2\Bigl(\frac{t}{4}\Bigr) \rho_S' \otimes E_1 -2
%\cos\Bigl(\frac{t}{4}\Bigr) \sin\Bigl(\frac{t}{4}\Bigr) [\sigma_x,
%\rho_S'] \otimes E_1 -4 \sin^2\Bigl(\frac{t}{4}\Bigr) \sigma_x \rho_S' \sigma_x \otimes E_1
&& = \left( e^{- \frac{t}{2} \sigma_x} \rho_S' e^{ \frac{t}{2} \sigma_x }\right)
\otimes E_1=X \rho_S' X^{\dagger} \otimes E_1,
\end{eqnarray}
where $X$ denotes  a generic transformation  of the form
$e^{\sigma_x \alpha}$ for $ \alpha $ real. Using (\ref{calcolo3}) with
$\rho_S'=Z_1 \rho_S Z_1^\dagger$, we have
\begin{eqnarray}\label{Calcolo4}
T \rho_S \otimes E_1 T^{\dagger} &=&
\left( Z_2 X Z_1 \rho_S Z_1^\dagger X^\dagger Z_2^\dagger \right)
\otimes E_1, \nonumber \\
T &:=& Z_2 \otimes {\bf 1} \, e^{i t \sigma_x \otimes \sigma_z} Z_1 \otimes {\bf 1},
\end{eqnarray}
and taking the partial trace, we obtain $Z_2 X
Z_1 \rho_S Z_1^\dagger X^\dagger Z_2^\dagger$. The claim follows
because, from Euler decomposition of $SU(2)$, every element of
$SU(2)$ can be written in the form $Z_2 X Z_1$. \epr

We remark that, given the structure of the Lie algebra ${\cal L}$,
we could have chosen a specific  (arbitrary) pure state, and the
proof of the previous proposition would have gone through.

\subsection{Full Indirect Controllability}

We conclude our investigation of the indirect controllability of a
system of two qubits studying Full Indirect Controllability. In
general, complete controllability does not imply  strong $FIC$ since
the state $\rho_S\otimes \rho_A=\frac{1}{4}({\bf 1} \otimes {\bf
1})$ is a fixed state for the total unitary dynamics. If we consider
the set $\Lambda_A$ to be the set of pure states for $A$,  we have
however the following result.

\bt{CCimpliesFIC} Assume the system $S+A$ is CC. Then $S$ is pure
FIC. \et \bpr We write the initial state of $S$ as $\rho_S :=
\sum_{j=1,2} r_j \vert j \rangle \langle j \vert$, with $\{\vert j
\rangle \}$ an orthonormal basis in ${\cal S}$, $\sum_{j=1,2} r_j =
1$, and the initial state of $A$ as $\rho_A : = \vert \psi \rangle
\langle \psi \vert$. Therefore, \be{raf1}\rho_S \otimes \rho_A =
\sum_j r_j \vert \psi_j \rangle \langle \psi_j \vert, \ee where
$\vert \psi_j \rangle = \vert j \rangle \otimes \vert \psi \rangle$,
and $\langle \psi_j \vert \psi_k \rangle = \delta_{jk}$. An
arbitrary unitary transformation of the composite system, denoted by
$U$, will act as \be{raf2}U \rho_S \otimes \rho_A U^{\dagger} =
\sum_{j=1,2} r_j \vert \phi_j \rangle \langle \phi_j \vert, \ee with
$\langle \phi_j \vert \phi_k \rangle = \delta_{jk}$, and $\vert
\phi_j \rangle$ arbitrary orthonormal vectors vectors in ${\mathcal
S} \otimes {\mathcal A}$. Now, consider the following particular
choices of $U$:

\noindent 1. $U$ is a unitary operator such that every vector $\vert
\phi_j \rangle$ is maximally entangled. This is always possible,
because there are 4 maximally orthonormal entangled vectors  in
${\mathcal S} \otimes {\mathcal A}$ (cf. \cite{NC}), and we need
only 2 of them. In this case, \be{raf3}\rho_S^{\prime} =
\sum_{j=1,2} r_j {\rm Tr}_A (\vert \phi_j \rangle \langle \phi_j \vert) =
\frac{1}{2} \, {\bf 1}, \ee that is, the initial $\rho_S$ is mapped
to the maximally mixed state;

\noindent 2. $U$ is the SWAP operator, that is $U \rho_S \otimes
\rho_A U = \rho_A \otimes \rho_S$. In this case, the initial
$\rho_S$ is mapped to the pure state $\vert \psi \rangle \langle
\psi \vert$.

Since the unitary group is path-connected, there is certainly a
continuous path in the space of states of the system $S$, connecting
the maximally mixed state to $\vert \psi \rangle \langle \psi
\vert$, and representing reachable states. By further using local
unitary operations acting on $S$, every final state can be obtained.
\epr

\section{CONCLUSIONS}
\label{conclusioni}

In this paper, we have formally introduced  the problem of indirect
controllability for quantum systems and we have given tools and
tests to describe the reachable set. In particular, we have
introduced  physically motivated notions and proved the general Theorem
\ref{gennegat} to conclude lack of indirect controllability. This
theorem is a test  at the Lie algebra level and therefore its
application implies only operations from linear algebra. The case
where both the target system and the accessor system are two level
systems is the simplest non trivial case which is also of great
importance in applications. To treat this system  we have made full
use of techniques in the Lie algebraic approach to quantum control.
Our first step has been to describe the possible dynamical Lie
algebras that can arise (Theorems \ref{gendynLiealg} and
\ref{OMS0}). These results are of interest on their own, and
generalize previous results in the literature.  With this
information at hand, we have proven in Theorem \ref{UICequiv} that we
need complete controllability of the whole system in order to have
unitary indirect controllability given any state of the accessor
system. However, if we are allowed to consider only pure states for the probe, we can
have unitary indirect controllability without complete
controllability (Proposition \ref{pure9}). When the state of the
accessor system is pure and we have complete controllability, we can
have full controllability on the target system, i.e., move the state
between arbitrary density matrices (Theorem \ref{CCimpliesFIC}).

This paper is a first step in the study of indirect controllability
and several questions are of interest for future research. Among
these, the extension of our results to systems with general
dimensions and an analysis of how indirect controllability of the
system $S$ through $A$ depends on the relative dimensions of $S$ and
$A$. In this work we have given tools but also indicated the
difficulties in pursuing this study including the fact that much of
the criteria and results have to be necessarily dependent on the
coordinates. We believe that a closer connection with the theory of
entanglement in quantum information would be fruitful in further
developing this important topic.

%%%%%%%%%%%%%%%%%%%%%%%%%%%%%%%%%%%%%%%%%%%%%%%%%%%%%%%%%%%%%%%%%%%%%%%%%%%%%%%%%%%%%%%%%%

\section{Expression of $C_0$, $C_x$, $C_y$ and
$C_z$ in example \ref{contoesempio}}
\noindent Recall that $\alpha_1$ is defined in (\ref{esa}). We have
\begin{eqnarray}\label{add1}
e^{\sigma_z \otimes {\bf 1}t_4} &=& c_2 {\bf 1} \otimes {\bf 1} + i s_2
\tilde \sigma_z \otimes {\bf 1}, \quad \alpha_2 := \frac{t_4}{2}, \\
e^{{\bf 1} \otimes \sigma_x a_1} &=& c_3 {\bf 1} \otimes {\bf 1} + i s_3
{\bf 1} \otimes \tilde \sigma_x, \quad \alpha_3 := \frac{a_1}{2}, \nonumber \\
e^{i \sigma_x  \otimes \sigma_x a_2} &=& c_4 {\bf 1} \otimes {\bf 1} + i s_4
\tilde \sigma_x  \otimes \tilde \sigma_x, \quad \alpha_4 := - \frac{a_2}{4}, \nonumber \\
e^{{\sigma_z} \otimes {\bf 1} s_1} &=& c_5 {\bf 1} \otimes {\bf 1} + i s_5
\tilde{\sigma_z} \otimes {\bf 1}, \quad \alpha_5 := \frac{s_1}{2}, \nonumber \\
e^{i \sigma_x  \otimes \sigma_z s_2} &=& c_6 {\bf 1} \otimes {\bf 1} + i s_6
\tilde \sigma_x  \otimes \tilde \sigma_z, \quad \alpha_6 := - \frac{s_2}{4},  \nonumber
\end{eqnarray}
having defined $c_j:=\cos({\alpha_j})$, $s_j:=\sin(\alpha_j)$ for $j=3,4$.
We have for  the matrices $C_0$, $C_x$, $C_y$ and $C_z$ in example
\ref{contoesempio}:
\begin{eqnarray}\label{C0}
C_0 &=& c_3 c_4 c_{2+5} c_{1+6} {\bf 1}- s_3 s_4 c_{2-5} c_{1+6} \tilde \sigma_x \nonumber \\
&& - s_3 s_4 s_{2-5} c_{1-6} \tilde
\sigma_y + i c_3 c_4 s_{2+5} c_{1-6} \tilde \sigma_z, \nonumber \\
C_x &=& i s_3 c_4 c_{2+5} c_{1-6} {\bf 1} + i c_3 s_4 c_{2-5} c_{1-6} \tilde \sigma_x \nonumber \\
&& i c_3 s_4 s_{2-5} c_{1+6} \tilde \sigma_y - s_3 c_4 s_{2+5} c_{1+6} \tilde \sigma_z, \nonumber \\
C_y &=& ic_3 s_4 c_{2-5} s_{1-6} {\bf 1} + i s_3 c_4 c_{2+5} s_{1-6} \tilde \sigma_x \nonumber \\
&& -i s_3 c_4 s_{2+5} s_{1+6} \tilde \sigma_y + c_3 s_4 s_{2-5} s_{1+6} \tilde \sigma_z, \nonumber \\
C_z &=& -i s_3 s_4 c_{2-5} s_{1+6} {\bf 1} + i c_3 c_4 c_{2+5} s_{1+6} \tilde \sigma_x \nonumber \\
&& -i c_3 c_4 s_{2+5} s_{1-6} \tilde \sigma_y - s_3 s_4 s_{2-5} s_{1-6} \tilde \sigma_z,
\end{eqnarray}
where we have defined $c_{i \pm j} := \cos (\alpha_i \pm \alpha_j)$, and $s_{i \pm j} := \sin (\alpha_i \pm \alpha_j)$.

\section{Proof of controllability in the case (\ref{calL2p})}
Using (\ref{cond2b}), we define
$\cos(\theta):=\frac{\alpha}{\sqrt{k}}$ and $\sin(\theta)
:=\frac{2\omega_A}{\sqrt{k}}$. For brevity we denote by $c$ and $s$,
$\cos(\theta)$ and $\sin(\theta)$ respectively. The goal is to show
that the Lie algebra generated by the matrices $\{ \Gamma_{x,y,z}^-,
\Gamma_y^+\}$ and the matrix $P:=i\sigma_c \otimes \sigma_z$ is the
full Lie algebra ${\mathfrak{su}}(4)$. This is the Lie algebra
generated by the matrices

\be{Ap1} \Gamma_x^-:= i \sigma_y \otimes \sigma_x -c i \sigma_x
\otimes \sigma_y + \frac{s}{2} \, {\bf 1} \otimes \sigma_x, \ee
\be{Ap2} \Gamma_z^-:= i \sigma_y \otimes \sigma_y + ci \sigma_x
\otimes \sigma_x + \frac{s}{2} \, {\bf 1} \otimes \sigma_y,  \ee
\be{Ap3} Z:= {\bf 1} \otimes \sigma_z,  \ee \be{Ap4} A:=c \sigma_z
\otimes {\bf 1} -2s i \sigma_y \otimes {\sigma_z}, \ee \be{Ap5} P:=i
\sigma_c \otimes \sigma_z, \ee with the first four matrices forming
a subalgebra. Applying the standard algorithm to calculate the Lie
algebra generated by a set of matrices (see,  e.g., \cite{Mikobook})
we first calculate all the Lie brackets of depth 1. We however do
not report Lie brackets which give zero and Lie brackets among the
first terms of (\ref{Ap1})-(\ref{Ap4}) since they do not give any
new directions. Moreover we also skip the Lie bracket with
$\Gamma_x^-$. Since $\Gamma_x^-$ is the Lie bracket of a linear
combination of $A$ and $Z$, say $L$, with $\Gamma_z^-$, from the
Jacobi identity, for any matrix $M$ we have
$[M,\Gamma_x^-]=-[L,[\Gamma_z^-,M]]-[\Gamma_z^-,[M,L]]$. So it is
enough to consider Lie brackets with $Z$, $A$ and $\Gamma_z^-$ only.
We calculate
\be{Q1}
    Q_1 := [P,A] = ic (y \sigma_x \otimes \sigma_z - x
\sigma_y \otimes \sigma_z) + \frac{s}{2} \, (z \sigma_x \otimes {\bf 1}
- x \sigma_z \otimes {\bf 1}),
\ee
\be{Q2}
Q_2:=4[P,\Gamma_z^-]=-y {\bf 1} \otimes \sigma_x +cx {\bf 1} \otimes
\sigma_y -2is \sigma_c \otimes \sigma_x.  \ee
 At step $2$ we take the Lie brackets obtained at step $1$ ($Q_1$ and $Q_2$) with the
 generating matrices (\ref{Ap2})-(\ref{Ap5}). We scale and eliminate
 directions that are already achieved. From $[Q_1,P]$, since $z\ne 0$ from C1', we obtain
 \be{R1}
R_1=c \sigma_c \otimes {\bf 1}.
 \ee
From $[Q_1,A]$, we obtain
\begin{equation}\label{R2}
R_2 := ic^2 z \sigma_z \otimes
\sigma_z + i s^2 y \sigma_y \otimes \sigma_z -\frac{sc}{2} \,
(y \sigma_z \otimes {\bf 1} + z \sigma_y \otimes {\bf 1}).
\end{equation}
From $[Q_1, \Gamma_z^-]$, we obtain
\begin{eqnarray}\label{R3} R_3 &:=&
\frac{c^2 y}{4} \, {\bf 1} \otimes \sigma_y - i \frac{s c}{2}y \sigma_x
\otimes \sigma_x + \frac{c x}{4} \, {\bf 1} \otimes \sigma_x + \nonumber \\
&& + \, i \frac{s}{2} \, (z \sigma_z \otimes \sigma_y + x \sigma_x
\otimes \sigma_y).
\end{eqnarray}
From $[Q_2,P]$ we obtain \be{R4} R_4 := i y
\sigma_c \otimes \sigma_y + ic x \sigma_c \otimes \sigma_x +
\frac{s}{2} \, {\bf 1} \otimes \sigma_y. \ee
% From $[Q_2, A]$ we obtain \be{R5}R_5:= 2iscx \sigma_y \otimes \sigma_x -2iscy
% \sigma_x \otimes \sigma_x - 4s^2 y {\bf 1} \otimes \sigma_y. \ee
% Actually this is wrong: the correct expression is [Q_2, A] = -2ys\Gamma_z^-
% without new directions
From $[Q_2, Z]$, we obtain \be{R5} R_5:= y {\bf 1} \otimes \sigma_y +cx {\bf 1} \otimes
\sigma_x + 2is \sigma_c \otimes \sigma_y.  \ee From $[Q_2,
\Gamma_z^-]$, we obtain \be{R6} R_6 := - ic^2 x \sigma_x \otimes
\sigma_z -i y \sigma_y \otimes \sigma_z + \frac{sc}{2} \, (y \sigma_z
\otimes {\bf 1} - z \sigma_y \otimes {\bf 1}).  \ee
Now calculate \be{S1} S_1 := [R_4,Z] = i y \sigma_c \otimes \sigma_x -icx
\sigma_c \otimes \sigma_y + \frac{s}{2} \, {\bf 1} \otimes \sigma_x. \ee
We have
\begin{equation}\label{Calcol78}
2 s c x y S_1 - 2 s y^2 R_4 + (c^2 x^2y + y^3) R_5 =
\end{equation}
$y^2(y^2 + c^2 x^2 - s^2) {\bf 1} \otimes \sigma_y + c x y (c^2
x^2 + y^2 + s^2) {\bf 1} \otimes \sigma_x.$
Now, under the
given assumption, the second term in the above expression is equal
to zero if and only if $cxy=0$. In fact the coefficient in
parentheses $(c^2 x^2+y^2+s^2)$ is not zero because this
would imply $s=0$, $x=0$ (since $c^2=1$) and $y=0$, contradicting
assumption C2'. Therefore formula (\ref{Calcol78}) gives the needed
extra local term unless at least one between $c$, $x$ and $y$ is
zero. We only need to consider these cases.

\vs

We notice that $c\ne 0$, otherwise C1' is violated. Moreover, if $x = y = 0$,
 we have $z = 1$, and $s \ne 0$ to fulfill C2'; in this case it is sufficient
to consider $[P, Q_2] = -\frac{s}{2} {\bf 1}\otimes \sigma_y$, which
is the extra desired local term. It remains to consider the cases
when $x = 0$ and $y \ne 0$, or viceversa. If $x = 0$,
(\ref{Calcol78}) gives the needed extra local term unless $y = \pm
s$. In this case, the local term ${\bf 1}\otimes \sigma_z$ is
obtained from $[Q_2, R_4]$ by discarding the component along $P$.
Finally, when $y = 0$ and $x \ne 0$, it is always possible to obtain
the local term ${\bf 1}\otimes \sigma_y$ by linearly combining $Q_2$
and $R_4$, since $c, s$ and $x$ are all real numbers.

%If $c=0$ then $s=1$ and we get the extra local term from $R_5$
%unless $y$ is also zero. However if $y$ is also zero $R_3$ would
%give $i\sigma_c \otimes \sigma_y$ (recall that
%$\sigma_c:=x\sigma_x+y\sigma_y+z\sigma_z$) and taking the Lie
%bracket of this with $i \sigma_c \otimes \sigma_z$ gives a term
%proportional to ${\bf 1}\otimes \sigma_x$ which is the extra desired
%local term. Therefore, the case $c=0$ is excluded.

% Assume now $y=0$. If $x$ is also zero $R_4$ gives the desired local
% unless $s=0$. However $s=0$ is the case treated in the first
% version  of the paper and therefore should be all right. In view of
% this fact we can always assume $s\not=0$. If $x\not=0$ $R_5$ gives
% $i\sigma_y \otimes \sigma_z$ and Lie bracket with $i \sigma_c
% \otimes \sigma_z$ gives the desired local.

% It remains to consider the case $x=0$ and $y$ and $c$ both
% different from zero. Assume this to be the case. Taking the Lie bracket of $R_4$ with
% $i\sigma_c \otimes \sigma_z$ and combining with $Q_2$ we obtain the
% extra desired local element.

%%%%%%%%%%%%%%%%%%%%%%%%%%%%%%%%%%%%%%%%%%%%%%%%%%%%%%%%%%%%%%%%%%%%%%%%%%%%%%%%%%%%%%%%

%\section*{Acknowledgment}
%
%D. D'Alessandro research was supported by NSF under Grant No.
%ECCS0824085. R. Romano acknowledges financial support from the Parisi Foundation.

%%%%%%%%%%%%%%%%%%%%%%%%%%%%%%%%%%%%%%%%%%%%%%%%%%%%%%%%%%%%%%%%%%%%%%%%%%%%%%%%

%%%%%%%%%%%%%%%%%%%%%%%%%%%%%%%%%%%%%%%%%%%%%%%%%%%%%%%%%%%%%%%%%%%%%%%%%%%%%%%%%%%%%%%%%

\end{document}